\documentclass[aps,prm,amsmath,amssymb,reprint,superscriptaddress,nofootinbib]{revtex4-2}
\usepackage{charter,graphicx,verbatim,threeparttable,float,amssymb,gensymb }
\usepackage{upgreek}
\usepackage{booktabs}
\begin{document}

\preprint{APS/123-QED}

\title{
Intricate magnetic landscape in antiferromagnetic kagome metal TbTi$_3$Bi$_4$ and interplay with \textit{Ln}$_{2-x}$Ti$_{6+x}$Bi$_9$ (\textit{Ln}: Tb$\cdots$Lu) shurikagome metals}

\author{Brenden R. Ortiz}
\email{ortizbr@ornl.gov}
\affiliation{Materials Science and Technology Division, Oak Ridge National Laboratory, Oak Ridge, TN 37831, USA}

\author{Heda Zhang} 
\affiliation{Materials Science and Technology Division, Oak Ridge National Laboratory, Oak Ridge, TN 37831, USA}

\author{Karolina G\'{o}rnicka} 
\affiliation{Materials Science and Technology Division, Oak Ridge National Laboratory, Oak Ridge, TN 37831, USA}

\author{David S. Parker}
\affiliation{Materials Science and Technology Division, Oak Ridge National Laboratory, Oak Ridge, TN 37831, USA}

\author{German D. Samolyuk}
\affiliation{Materials Science and Technology Division, Oak Ridge National Laboratory, Oak Ridge, TN 37831, USA}

\author{Fazhi Yang} 
\affiliation{Materials Science and Technology Division, Oak Ridge National Laboratory, Oak Ridge, TN 37831, USA}

\author{Hu Miao} 
\affiliation{Materials Science and Technology Division, Oak Ridge National Laboratory, Oak Ridge, TN 37831, USA}

\author{Qiangsheng Lu} 
\affiliation{Materials Science and Technology Division, Oak Ridge National Laboratory, Oak Ridge, TN 37831, USA}

\author{Robert G. Moore} 
\affiliation{Materials Science and Technology Division, Oak Ridge National Laboratory, Oak Ridge, TN 37831, USA}

\author{Andrew F. May} 
\affiliation{Materials Science and Technology Division, Oak Ridge National Laboratory, Oak Ridge, TN 37831, USA}

\author{Michael A. McGuire} 
\affiliation{Materials Science and Technology Division, Oak Ridge National Laboratory, Oak Ridge, TN 37831, USA}

\date{\today}
\begin{abstract}
Here we present the discovery and characterization of the kagome metal TbTi$_3$Bi$_4$ in tandem with a new series of compounds, the \textit{Ln}$_{2-x}$Ti$_{6+x}$Bi$_9$ (\textit{Ln}: Tb--Lu) shurikagome metals.
We previously reported on the growth of the \textit{Ln}Ti$_3$Bi$_4$ (\textit{Ln}: La--Gd$^{3+}$, Eu$^{2+}$, Yb$^{2+}$) family, a chemically diverse and exfoliable series of kagome metals with complex and highly anisotropic magnetism.
However, unlike the La--Gd analogs, TbTi$_3$Bi$_4$ cannot be synthesized by our previous methodology due to phase competition with \textit{Ln}$_{2-x}$Ti$_{6+x}$Bi$_9$  (x$\sim$1.7--1.2). 
Here we discuss the phase competition between the \textit{Ln}Ti$_3$Bi$_4$ and \textit{Ln}$_{2-x}$Ti$_{6+x}$Bi$_9$ families, helping to frame the difficulty in synthesizing \textit{Ln}Ti$_3$Bi$_4$ compounds with small \textit{Ln} species and providing a strategy to circumvent formation of \textit{Ln}$_{2-x}$Ti$_{6+x}$Bi$_9$.
Detailed characterization of the magnetic and electronic transport properties on single crystals of TbTi$_3$Bi$_4$ reveals a highly complex landscape of magnetic phases arising from an antiferromagnetic ground state. 
A series of metamagnetic transitions creates at least 5 unique magnetic phase pockets, including a 1/3 and 2/3 magnetization plateau. Further, the system exhibits an intimate connection between the magnetism and magnetotransport, exhibiting sharp switching from positive (+40\%) to negative magnetoresistance (--50\%). 
Like the \textit{Ln}Ti$_3$Bi$_4$ kagome metals, the \textit{Ln}$_{2-x}$Ti$_{6+x}$Bi$_9$ family exhibits quasi-2D networks of titanium and chains of rare-earth. We present the structures and some basic magnetic properties of the \textit{Ln}$_{2-x}$Ti$_{6+x}$Bi$_9$ family alongside our characterization of the newly discovered TbTi$_3$Bi$_4$. 
\footnote{Notice: This manuscript has been authored by UT-Battelle, LLC, under contract DE-AC05-00OR22725 with the US Department of Energy (DOE). The US government retains and the publisher, by accepting the article for publication, acknowledges that the US government retains a nonexclusive, paid-up, irrevocable, worldwide license to publish or reproduce the published form of this manuscript, or allow others to do so, for US government purposes. DOE will provide public access to these results of federally sponsored research in accordance with the DOE Public Access Plan (https://www.energy.gov/doe-public-access-plan).} 

\end{abstract}
\maketitle
\section{Introduction}

Spurred on by continued discoveries in the \textit{A}\textit{M}$_3$\textit{X}$_5$, \textit{A}\textit{M}$_6$\textit{X}$_6$, and CoSn-type compounds, kagome metals have proven to be a rich space for exploring the intersection of solid-state chemistry and fundamental condensed matter physics. The impetus for studying kagome metals is derived from the basic electronic structure of a kagome toy model -- which creates the potential for Dirac points, flat bands, and Van Hove singularities.~\cite{park2021electronic,PhysRevB.87.115135,kiesel2013unconventional,meier2020flat}. Indeed, the latent potential within the kagome motif is often unlocked by designing materials where the Fermi level (E$_\text{F}$) is aligned with these electronic features, as exemplified by the \textit{A}V$_3$Sb$_5$ (\textit{A}: K, Rb, Cs) kagome materials. In \textit{A}V$_3$Sb$_5$ compounds, the Fermi level is intrinsically tuned to a Van Hove singularity, causing these nonmagnetic materials to be heavily investigated for complex interactions between electronic and structural instabilities (charge density wave, potential unconventional superconductivity, non-trivial topology).\cite{wilsonReview,ortizCsV3Sb5,ortiz2020KV3Sb5,RbV3Sb5SC,ortiz2021fermi,zhao2021cascade,hu2022coexistence,kang2022microscopic,jiang2021unconventional} The recent discovery of CsCr$_3$Sb$_5$ has also included the possibility of exploring the interaction between a possible spin density wave and superconductivity in proximity to a flat band near E$_\text{F}$.\cite{liu2023superconductivity}

Indeed, CsCr$_3$Sb$_5$ exemplifies another key direction for the exploration of new kagome materials -- the interaction between magnetic sublattices and the underlying kagome band structure. The development of new magnetic kagome systems (either with magnetic kagome sublattices, or magnetic supporting sublattices) has been remarkable. Consider, for example, the chemical diversity of the CoSn-family and it's grandchildren, the \textit{A}\textit{M}$_6$\textit{X}$_6$ compounds.\cite{wang2022electronic, PhysRevLett.127.266401,PhysRevB.103.014416,PhysRevB.104.235139,PhysRevMaterials104202,PhysRevB.103.014416,PhysRevB.106.115139,sciadv_abe2680,PhysRevLett.129.216402,Yin_2020,PhysRevMaterials.6.105001,PhysRevMaterials.6.083401,ZhangShao-ying_2001,PhysRevLett.126.246602} The structural relationship between CoSn and the \textit{A}\textit{M}$_6$\textit{X}$_6$ is an instructive one, as the two compounds are fundamentally linked. The CoSn prototype features large interstitial voids between the hexagonal plaquettes of adjacent kagome networks, which offers a unique opportunity for generating new kagome materials. By adding an electropositive filler atom in these natural voids, we can produce ordered superlattices -- giving birth to the \textit{A}\textit{M}$_6$\textit{X}$_6$ (HfFe$_6$Sn$_6$, \textit{P}6/\textit{mmm}) prototype.\cite{fredrickson2008origins} The more complex ordering of the \textit{A}\textit{M}$_6$\textit{X}$_6$ creates an explosion of structural diversity, as the ordered superlattice exhibits multiple stacking variations and intergrowths like ErFe$_6$Sn$_6$ (\textit{Amam}), DyFe$_6$Sn$_6$ (\textit{Cmmm}), and YFe$_6$Sn$_6$ (\textit{Immm}).\cite{el1991crystal} The titular series -- the \textit{A}\textit{M}$_3$\textit{X}$_4$ kagome metals -- also contain remnants of the \textit{A}\textit{M}$_6$\textit{X}$_6$ stacking, producing a prototypical grandchild of the CoSn family.

We previously reported the discovery of both the antimonide \textit{Ln}V$_3$Sb$_4$ (\textit{Ln}: Eu$^{2+}$, Yb$^{2+}$)
\cite{ortiz2023ybv} and bismuthide \textit{Ln}Ti$_3$Bi$_4$ (\textit{Ln}: La...Gd$^{3+}$, Eu$^{2+}$, Yb$^{2+}$) kagome metals,\cite{ortiz2023evolution} which make up a larger family of \textit{A}\textit{M}$_3$\textit{X}$_4$ compounds.\cite{ortiz2019new,ortiz2023evolution,ovchinnikov2018synthesis,ovchinnikov2019bismuth,motoyama2018magnetic,chen2023134,guo2023134} This family boasts both magnetic and non-magnetic compounds through the diverse choice of \textit{A} cations, which are typically electropositive rare-earth and alkaline-earth metals. The \textit{A}-site elements comprise zig-zag chains which run parallel to the \textit{M}-site kagome layers. Remarkably, unlike the CoSn and \textit{A}\textit{M}$_6$\textit{X}$_6$ parent structures, the \textit{A}\textit{M}$_3$\textit{X}$_4$ compounds exhibit a much more quasi-2D, layered, exfoliable crystal structure. Developing quasi-2D kagome platforms is critical, as the dimensionality of the kagome network and the supporting sublattices have a profound effect on the realization of key features in the electronic structure.\cite{jovanovic2022simple} Crystals are easily exfoliated (similar to the \textit{A}\textit{M}$_3$\textit{X}$_5$ family), which makes them an attractive option for exploring both magnetism and the kagome band structure. Accordingly, the \textit{A}\textit{M}$_3$\textit{X}$_4$ compounds have been the focus of a growing body of work, particularly studies involving angle-resolved photoemission spectroscopy (ARPES).\cite{chen2023134,chen2023sm134,hu2023magnetic,mondal2023observation,jiang2023direct,sakhya2023observation}

Under the conditions presented in our previous work, \textit{Ln}Ti$_3$Bi$_4$ compounds containing rare-earth atoms smaller than Gd$^{3+}$ did not appear. In this work, we present a modified growth procedure which allows stabilization of TbTi$_3$Bi$_4$, the smallest rare-earth cation reported in the \textit{Ln}Ti$_3$Bi$_4$ structure thus far. Our work suggests that TbTi$_3$Bi$_4$ is a relatively low temperature phase ($\sim$500\degree C) and is suppressed through competition with a new series of compounds, the \textit{Ln}$_{2-x}$Ti$_{6+x}$Bi$_9$ (\textit{Ln}: Tb--Lu) shurikagome metals. By quenching fluxes from 1100\degree C to 600\degree C, we suppress the growth of Tb$_{2-x}$Ti$_{6+x}$Bi$_9$ and encourage the resorption of any shurikagome nuclei. A subsequent cool from 600\degree C to 400\degree C encourages the growth of single crystals of TbTi$_3$Bi$_4$. This work details our rationale for the stabilization of TbTi$_3$Bi$_4$ alongside a discussion of its magnetic and electronic transport properties. Remarkably, TbTi$_3$Bi$_4$ presents some of the most rich and complex magnetism in the known \textit{A}\textit{M}$_3$\textit{X}$_4$ compounds, reinforcing the need to understand and control the formation of new phases. Our work also briefly investigates the structural and magnetic properties of the new \textit{Ln}$_{2-x}$Ti$_{6+x}$Bi$_9$ (\textit{Ln}: Tb--Lu) metals. 

\section{Methods}
\subsection{Single Crystal Synthesis}
TbTi$_3$Bi$_4$ single crystals are grown through a bismuth self-flux. Tb metal (Ames), Ti powder (Alfa 99.9\%), and Bi shot (Alfa 99.999\% low-oxide) were placed into a 5~mL Canfield crucibles fitted with a catch crucible and a porous frit\cite{canfield2016use} at a ratio of 2:3:20 Tb:Ti:Bi. The crucibles were sealed under approximately 0.7~atm of argon gas in fused silica ampoules. Samples were subsequently heated to 1100\degree C at a rate of 200\degree C/hr and thermalized at 1100\degree C for 12~h. The samples were then quenched by removing them from the furnace and immediately placing them into a different furnace preheated to 600\degree C. The growth proceeded with a cooling rate of 1\degree C/hr to 400\degree C. Excess bismuth was removed through centrifugation at 400\degree C. 

\textit{Ln}$_{2-x}$Ti$_{6+x}$Bi$_9$ single crystals are grown through a bismuth self-flux as well. \textit{Ln} metal (Ames), Ti powder (Alfa 99.9\%), and Bi shot (Alfa 99.999\% low-oxide) were placed into a 5~mL Canfield crucibles fitted with a catch crucible and a porous frit\cite{canfield2016use} at a ratio of 2:3:20 \textit{Ln}:Ti:Bi. The crucibles were sealed under approximately 0.7~atm of argon gas in fused silica ampoules. \textit{Ln}$_{2-x}$Ti$_{6+x}$Bi$_9$ single crystals are stable over a wide range of fluxes (e.g., \textit{Ln}:Ti:Bi of 2:3:12, 2:3:20, 1:1:3). Crystals featured here were grown at a ratio of 2:3:12, though the composition of the crystals did not change appreciably under the various flux compositions investigated here. In particular, we are currently unable to grow the idealized, fully saturated \textit{Ln}$_{2}$Ti$_{6}$Bi$_9$. Fluxes were heated to 1100\degree C at a rate of 200\degree C/hr and thermalized at 1100\degree C for 12~h. The samples were then cooled linearly from 1100\degree C to 600\degree C at a rate of 2\degree C/hr. Excess bismuth was removed through centrifugation at 600\degree C. 

\subsection{Bulk Characterization}
Single crystals of both TbTi$_3$Bi$_4$ and \textit{Ln}$_{2-x}$Ti$_{6+x}$Bi$_9$ were mounted on kapton loops with Paratone oil for single crystal x-ray diffraction (SCXRD). Diffraction data were collected at 100~K on a Bruker D8 Advance Quest diffractometer with a graphite monochromator using Mo K$\alpha$ radiation ($\lambda$ = 0.71073~\AA). Data integration, reduction, and structure solution was performed using the Bruker APEX4 software package. As noted previously, diffraction on \textit{Ln}Ti$_3$Bi$_4$ crystals is challenging and requires face-indexing absorption corrections.\cite{ortiz2023evolution} All atoms were refined with anisotropic thermal parameters. CIF files for all structures are included in the ESI. A TF20 X-Ray System (LAUE-V-674-AC-GigE from Photonic Science) was used to characterize and orient single crystals for magnetotransport and magnetization measurements. Elemental analysis was carried out on as-grown crystals using a Hitachi-TM3000 microscope equipped with a Bruker Quantax 70 EDS system. 

Magnetization measurements (300~K -- 1.8~K) on crystals of TbTi$_3$Bi$_4$ and \textit{Ln}$_{2-x}$Ti$_{6+x}$Bi$_9$ were performed in a 7~T Quantum Design Magnetic Property Measurement System (MPMS3) SQUID magnetometer in vibrating-sample magnetometry (VSM) mode. Additional measurements on TbTi$_3$Bi$_4$ (1.8~K -- 0.3~K) utilized the Quantum Design iHe-3 He$^{3}$ insert for the MPMS3. AC susceptibility measurements (3.5~K -- 0.1~K) on Tm$_{2-x}$Ti$_{6+x}$Bi$_9$ utilized a Quantum Design Dilution Refrigerator insert in a Quantum Design 9~T Dynacool PPMS. The lowest frequency data (100~Hz) was normalized to the MPMS3 data from 4~K to 1.8~K to create a continuous data set from 300~K to 0.1~K. Angle-resolved magnetization measurements on TbTi$_3$Bi$_4$ were performed on a 7~T Quantum Design Magnetic Property Measurement System (MPMSXL) equipped with a rotator stage. 

A note regarding hysteresis in TbTi$_3$Bi$_4$; crystals show substantial magnetic hysteresis, and considerable care must be taken to account for magnetic and thermal history. While samples show negligible changes in low-field ($<$500~Oe) measurements under field-cooled (FC) or zero-field-cooled (ZFC) conditions, field loops up to 3~T must be performed with consideration for the ZFC (virgin magnetization) curve. Additionally, subsequent measurements under varying temperatures (e.g. M vs. H curves at different temperatures) need to heat up above the transition temperature to ``erase'' the magnetic history. Specific details for each magnetic and transport measurement are included in the text alongside the respective figures. 

Electronic transport measurements on TbTi$_3$Bi$_4$ were performed in a Quantum Design 9~T Dynacool PPMS. Single crystals were mounted to a sheet of sapphire, which was subsequently adhered to the sample puck stage using GE varish. The sapphire could be rotated between different orientations to achieve measurements with the magnetic field both within and out-of-plane. Sample surfaces were cleaned by removing several layers using adhesive tape, and then contacts were made using silver paint (DuPont cp4929N-100) and platinum wire (Alfa, 0.05~mm Premion 99.995\%) in a six-wire geometry for simultaneous measurements of the Hall effect and magnetoresistance. As alluded to in the magnetization methods, care must be taken to account for magnetic and thermal history. We are primarily focused on the magnetotransport results that are repeatable under continued field cycling, and will typically omit virgin magnetoresistance curves. In this case, we typically begin each measurement in a fully-polarized (e.g. 9~T) state. All transport measurements utilized a probe current of 8~mA.

Heat capacity measurements (300--1.8~K) on TbTi$_3$Bi$_4$, Tb$_{2-x}$Ti$_{6+x}$Bi$_9$, and Er$_{2-x}$Ti$_{6+x}$Bi$_9$ were performed in a Quantum Design 9~T Dynacool Physical Property Measurement System (PPMS). Additional heat capacity measurements (3.5--0.1~K) on TbTi$_3$Bi$_4$ utilized the Quantum Design Dilution Refrigerator insert for the Quantum Design 9~T Dynacool PPMS. Thermal contact was achieved using Apezion N-grease.

\subsection{Electronic Structure Calculations}

In an effort to describe the underlying electronic structure of the \textit{Ln}$_2$Ti$_6$Bi$_9$ family, we have performed first-principles all-electron density
functional theory calculations within the generalized gradient approximation, as in several
of our recent works.\cite{ortiz2023evolution,taddei2023zigzag,sales2022chemical,sala2022ferrimagnetic,hobbis2019structural} To avoid 4f-electron-related complications we have chosen to study the Lu-based end-member Lu$_2$Ti$_6$Bi$_9$ in a non-spin-polarized configuration, with Lu having a full 4f shell and thereby no magnetic character. We have chosen to neglect potential structural disorder by studying the fully-ordered, stoichiometric structure of Lu$_2$Ti$_6$Bi$_9$. The results should be generally applicable to other lanthanides in this family, presuming that relevant measurements (such as ARPES) are performed at temperatures well above the onset of any magnetic order.\cite{pokharel2018negative} This is as in our previous work.\cite{ortiz2023evolution}  The linearized augmented plane-wave (LAPW) basis was employed, with an RKmax (the product of the smallest muffin-tin sphere radius and largest planewave expansion wavevector) of 8.0 and sphere radii of 2.50~Bohr for Bi and Lu were chosen, and 2.41~Bohr for Ti. Neither spin-orbit coupling nor the GGA+U approach (i.e., a ‘U’) was included.

\section{Results and Discussion}
\subsection{Crystal Structure and Synthesis}

\begin{figure}
\includegraphics[width=\linewidth]{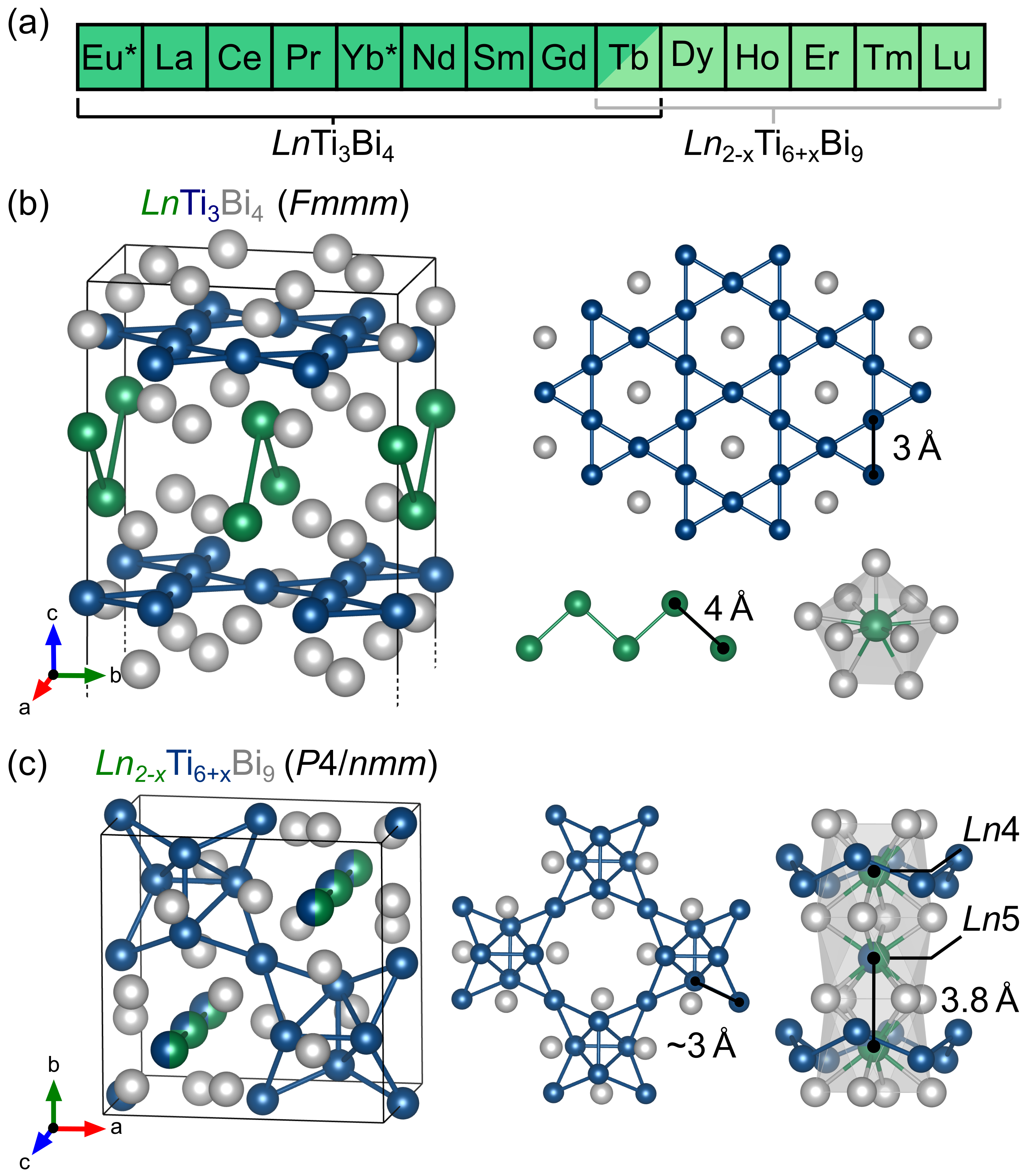}
\caption{(a) Simple schematic summarizing the phase stability of the \textit{Ln}Ti$_3$Bi$_4$ family and \textit{Ln}$_{2-x}$Ti$_{6+x}$Bi$_9$ across the lanthanide row. Elements are sorted by the 9-coordinate Shannon radius, where asterisks denote divalent species. (b) Crystal structure of \textit{Ln}Ti$_3$Bi$_4$, highlighting kagome network of Ti and \textit{Ln}--\textit{Ln} zig-zag chains. (c) Crystal structure of \textit{Ln}$_{2-x}$Ti$_{6+x}$Bi$_9$, highlighting the Ti ``shurikagome'' network and the mixed \textit{Ln}/Ti quasi-1D chains formed by the \textit{Ln}4 and \textit{Ln}5 sublattices. Coordination polyhedra for both structures are drawn for bonds less than $\sim$3.2\AA.}
\label{fig:1}
\end{figure}

Our prior work identified that the \textit{Ln}Ti$_3$Bi$_4$ (\textit{Ln}: La...Gd$^{3+}$, Eu$^{2+}$, Yb$^{2+}$) kagome metals are readily grown from a bismuth self-flux.\cite{ortiz2023evolution} However, growth of \textit{Ln}Ti$_3$Bi$_4$ crystals containing smaller rare-earth metals (\textit{Ln}: Tb--Lu) was stymied due to extensive crystallization of the \textit{Ln}$_{2-x}$Ti$_{6+x}$Bi$_9$ compounds.  Figure \ref{fig:1}(a) provides a simple schematic showing the lanthanide row sorted in order of the 9-coordinate Shannon ionic radius. Within the \textit{Ln}Ti$_3$Bi$_4$ compounds, both Eu and Yb appear to preferentially adopt the 2+ oxidation state. As such, the radii of Eu$^{2+}$ and Yb$^{2+}$ are analogous to the larger La$^{3+}$ and Pr$^{3+}$, respectively. Large cations preferentially stabilize the \textit{Ln}Ti$_3$Bi$_4$ phase, and small cations preferentially stabilize \textit{Ln}$_{2-x}$Ti$_{6+x}$Bi$_9$. Within this work, we demonstrate that Tb represents a unique cross-over between the two phases, and both TbTi$_3$Bi$_4$ and Tb$_{2-x}$Ti$_{6+x}$Bi$_9$ can be obtained by tuning the growth profiles.

Figure \ref{fig:1}(b,c) details the crystal structure of both the \textit{Ln}Ti$_3$Bi$_4$ (b) and \textit{Ln}$_{2-x}$Ti$_{6+x}$Bi$_9$ (c) compounds. Note that Figure \ref{fig:1}(b) presents only half of the \textit{Ln}Ti$_3$Bi$_4$ unit cell, with the omitted half generated by an inversion operation (space group \textit{Fmmm}). \textit{Ln}Ti$_3$Bi$_4$ contains two key structural motifs of interest; 1) a Ti--Ti kagome network, and 2) \textit{Ln}--\textit{Ln} zig-zag chains. The reduced symmetry of the unit cell is driven by the quasi-1D zig-zag chains, which necessitate a maximum of 2-fold rotational symmetry. Despite the orthorhombic structure, the kagome network is negligibly distorted ($<$0.1\AA~out-of-plane buckling). Accordingly, the orthorhombic \textit{b/a} lattice parameter ratio ($b/a\sim1.74$) is very nearly the nominal $b/a=\sqrt{3}\sim1.73$ applicable to the hexagonal symmetry. The zig-zag chains dominate the magnetic properties of these materials, with the nearest-neighbor intra-chain \textit{Ln}--\textit{Ln} distance ($\sim$4~\AA) substantially closer than the inter-chain distances ($\sim$6~\AA). We typically model the rare-earth local coordination as approximately 9-fold, though the coordination is not uniform and requires a small range of bond lengths (3.3~\AA -- 3.6~\AA) to recover the polyhedral environment shown in Figure \ref{fig:1}(b).

The \textit{Ln}$_{2-x}$Ti$_{6+x}$Bi$_9$ (\textit{P}4/\textit{nmm}) compounds can be understood as a heavily alloyed form of Ti$_8$Bi$_9$\cite{richter1997preparation}, where \textit{Ln} atoms preferentially occupy two of the Ti sites (there are 4 unique Ti sites in Ti$_8$Bi$_9$). If the substitution was perfect, (e.g. \textit{Ln}$_{2}$Ti$_{6}$Bi$_9$)  the \textit{Ln} atoms would create a nearly ideal 1D chain that cuts perpendicular to the Ti--Ti network. The \textit{Ln}--\textit{Ln} distance is closer in \textit{Ln}$_{2}$Ti$_{6}$Bi$_9$ ($\sim$3.8\AA) when compared to \textit{Ln}Ti$_3$Bi$_4$ ($\sim$4\AA), owing to the contraction from partial occupancy of Ti. The remaining Ti atoms form a unique network made of up Ti$_4$ tetrahedra interconnected with corner-sharing triangles. Paying homage to both the linguistic origin and the corner-sharing coordination of the kagome lattice, the tessellated unit resembles a four-pointed \textit{hira-shuriken} and we've taken to nicknaming these compounds ``shurikagome'' metals. 

Within Ti$_8$Bi$_9$, the two sites that are preferentially occupied by \textit{Ln} in \textit{Ln}$_{2}$Ti$_{6}$Bi$_9$ (\textit{Ln}4 and \textit{Ln}5 in Figure \ref{fig:1}(c)) possess considerably different coordination than the Ti that comprise the shurikagome network. Considering bonds $<$3.2~\AA, Ti atoms within the shurikagome network have an approximate coordination of 10 (4 Ti--Ti bonds, 6 Ti--Bi bonds) and a polyhedral volume of approximately 50.7~\AA$^3$. However, the \textit{Ln}4 and \textit{Ln}5 sites possess a coordination of 8, entirely from \textit{Ln}--Bi bonds, and form slightly distorted square anti-prisms. \textit{Ln}4 and \textit{Ln}5 are also distinct from each other. \textit{Ln}4 sits within the plane of the shurikagome network, and has substantially larger polyhedral volume (56.9~\AA$^3$) owing to the support of the Ti--Ti scaffolding. The \textit{Ln}5 site is slightly collapsed along \textit{c}, leading to a correspondingly lower volume (49.5~\AA$^3$) and a marginally more distorted coordination. As a conceptually pleasing observation, the rare-earth atoms appear to exclusively occupy the 8-coordinate sites, and preferentially occupy the largest of the mixed \textit{Ln}/Ti sites. From single crystal diffraction results, the larger site is approximately 75--90\% occupied by rare-earth, compared to 50--75\% for the smaller site. As such, the total stoichiometry can be more precisely written as \textit{Ln}$_{2-x}$Ti$_{6+x}$Bi$_9$ (x$\sim$1.7--1.2). CIF files and associated crystallograpic tables for both TbTi$_3$Bi$_4$ and the \textit{Ln}$_{2-x}$Ti$_{6+x}$Bi$_9$ compounds are included in the ESI. We will discuss trends in the \textit{Ln} substitution alongside the magnetic properties in a later section.

\begin{figure}
\includegraphics[width=\linewidth]{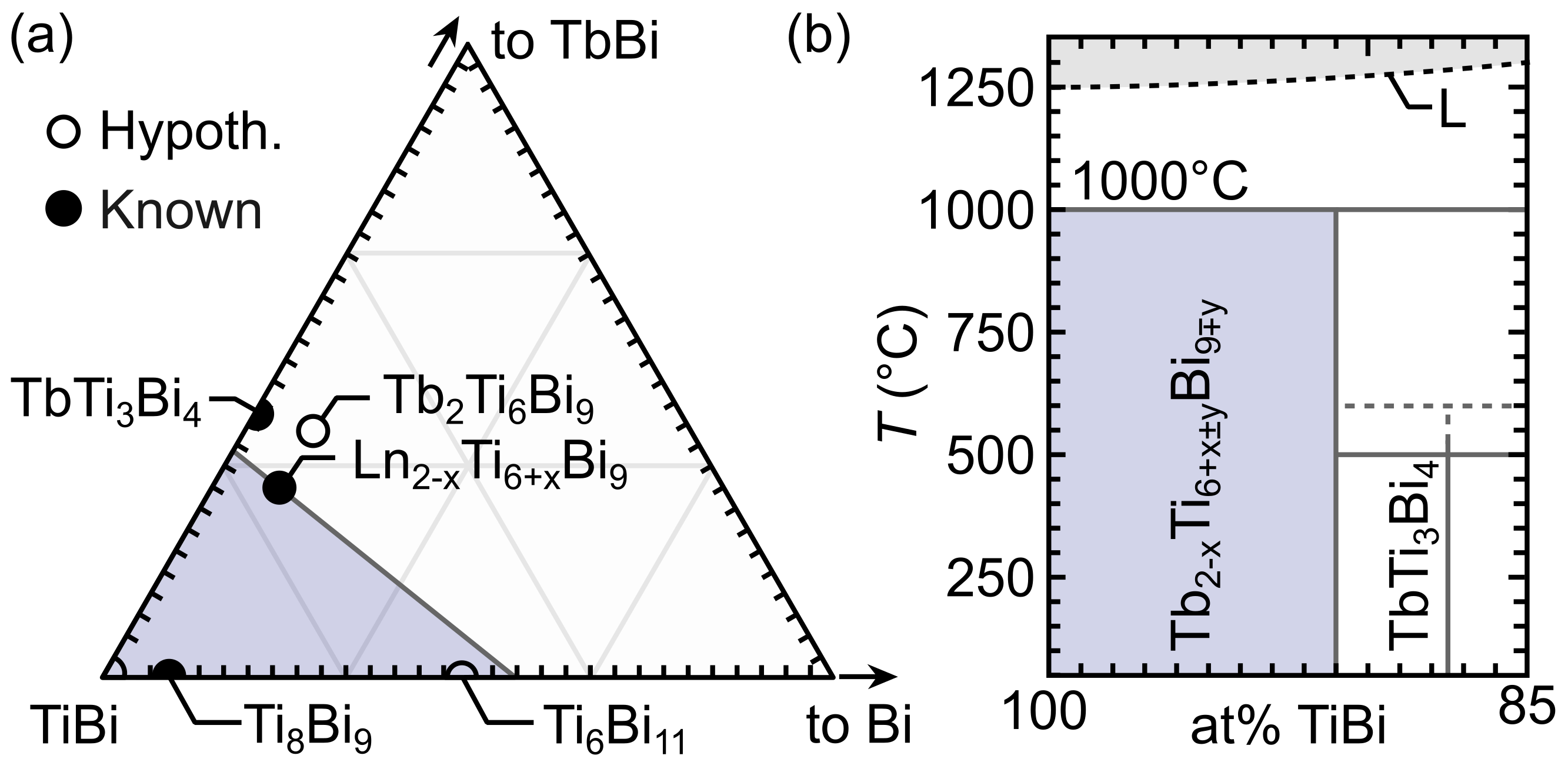}
\caption{(a) Small section of Tb--Ti--Bi phase diagram, highlighting the competing phases near TbTi$_3$Bi$_4$. The Alkemade triangles formed by TbBi--\textit{Ln}$_{2-x}$Ti$_{6+x}$Bi$_9$--Bi is stable above 600\degree C, but fragments into two triangles as the isotherm temperature is lowered and TbTi$_3$Bi$_4$ becomes stable. (b) An approximate projection of the phase diagram along the TbBi--TiBi pseudobinary line, highlighting key results of our crystal growth experiments. The estimated stability of TbTi$_3$Bi$_4$ is shown, with the phase stable to at least 500\degree C and possibly as high at 600\degree C, as indicated by the dashed line.}
\label{fig:2}
\end{figure}

As alluded to before, TbTi$_3$Bi$_4$ does not appear to form under conditions presented in our first manuscript.\cite{ortiz2023evolution} Samples of large rare-earth \textit{Ln}Ti$_3$Bi$_4$ compounds are readily grown through a bismuth self-flux cooled from 1100\degree C to 600\degree C. However, reactions utilizing Tb will exclusively form Tb$_{2-x}$Ti$_{6+x}$Bi$_9$. To help understand the shift towards crystallization of the Tb$_{2-x}$Ti$_{6+x}$Bi$_9$ phase, we first turned to examine the known phases in the vicinity of TbTi$_3$Bi$_4$. Figure \ref{fig:2}(a) shows a small section of the Tb--Ti--Bi phase diagram near several of the phases of interest. Data for the existing binary and pseudobinary lines, including the single phase region surrounding Ti$_8$Bi$_9$ were adapted from published Ti--Bi, Tb--Bi binary diagrams.\cite{vassilev2006contribution,okamoto1998bi} The large Ti--Bi offstoichiometry in samples of Ti$_8$Bi$_9$ is remarkable, and suggests that solid solubility exits beyond where the hypothetical compound Bi$_2$Ti$_{6}$Bi$_9$ exists (written in this form to highlight substitution of Bi on the mixed sites). There have also been reports that oxygen atoms can substitute in the center of the Ti$_4$ tetrahedra, creating compositions like Ti$_8$Bi$_9$O$_{0.25}$.\cite{yamane2018crystal} Such observations are consistent with the structural flexibility needed to host the \textit{Ln}$_\text{Ti}$ disorder proposed here. 

It appears that the fully-occupied Tb$_2$Ti$_6$Bi$_9$ is not stable under conditions explored here, as flux growths containing different concentrations of \textit{Ln}:Ti:Bi consistently produce samples of Tb$_{2-x}$Ti$_{6+x}$Bi$_9$ which are sub-occupied. This is fortunate from the perspective of TbTi$_3$Bi$_4$, as the presence of Tb$_2$Ti$_6$Bi$_9$ would hinder our ability to form TbTi$_3$Bi$_4$ from a bismuth-rich (low-temperature melting point) mixture. Given the absence of Tb$_2$Ti$_6$Bi$_9$, we hypothesized that TbTi$_3$Bi$_4$ could potentially be accessible from a Bi self-flux. However, if the phase exists, it must be a low-temperature phase, which is out-competed by the thermal stability of Tb$_{2-x}$Ti$_{6+x}$Bi$_9$. 

One preliminary observation that supported our hypothesis was the rounding (resorption) of the corners and edges of Tb$_{2-x}$Ti$_{6+x}$Bi$_9$ crystals at temperatures near 600\degree C (flux composition 2:3:20). Samples quenched at higher temperatures (e.g. 800\degree C) showed nicely faceted Tb$_{2-x}$Ti$_{6+x}$Bi$_9$ samples with no resorption, suggesting that the shurikagome is either: 1) unstable at low temperatures, or 2) is consumed by crystallization of a low-temperature phase. A similar phenomenon was noted with more Tb-rich compositions (flux composition 5:5:20) and the high melting point TbBi binary. Given sufficient time at low-temperatures, the decomposed corners of Tb$_{2-x}$Ti$_{6+x}$Bi$_9$ became encased in recrystallized agglomerates of TbTi$_3$Bi$_4$. This highlighted that TbTi$_3$Bi$_4$ is a low-temperature phase and evolves from the dissolution of the high-temperature compounds. Subsequent quenching experiments further suggested that TbTi$_3$Bi$_4$ is likely only stable below 600\degree C. 

Thus, we propose that TbTi$_3$Bi$_4$ is a low-temperature phase with a maximum stability of approximately 600\degree C in a bismuth self-flux. Upon cooling, it is preceded by either TbBi or Tb$_{2-x}$Ti$_{6+x}$Bi$_9$, depending on the flux composition. To circumvent formation of the competing phases, we quench samples from 1100\degree~C to 600\degree C. This step prevents extended nucleation and growth of Tb$_{2-x}$Ti$_{6+x}$Bi$_9$. The samples are then slow-cooled from 600\degree C to 400\degree C at a rate of 1\degree C/hr, ensuring resorption of any shurikagome nuclei and encouraging growth of TbTi$_3$Bi$_4$. Batches grown in this manner are primarily composed of nicely-faceted TbTi$_3$Bi$_4$. We are pleased to see that these observations agree nicely with recent results from other groups.\cite{cheng2024giant,guo20241}

Crystals of TbTi$_3$Bi$_4$ are 1--5~mm pseudo-hexagonal crystals with a brilliant metallic luster. They are extremely soft, readily exfoliated parallel to the \textit{ab} plane, and are lightly air-sensitive. In the antimonide analog, we have observed that both \textit{M}-Sb and \textit{Ln}-Sb terminations can be achieved through exfoliation, and we expect similar results for the bismuthide.\cite{ortiz2019new} Crystals of \textit{Ln}$_{2-x}$Ti$_{6+x}$Bi$_9$ are typically truncated square plates (nearly octagonal) with a dark gray luster. Crystals are substantially more brittle than the \textit{Ln}Ti$_3$Bi$_4$ compounds but can be readily cleaved parallel to the \textit{ab}-plane (perpendicular to the \textit{Ln} chains) with a sharp blade. \textit{Ln}$_{2-x}$Ti$_{6+x}$Bi$_9$ crystals are substantially more sensitive to air, and crystals will completely disintegrate within a few hours. Curiously, both samples decompose in the same manner. Upon extended exposure, the samples swell substantially and spall layer-by-layer.

\subsection{Properties of TbTi$_3$Bi$_4$}

\begin{figure*}
\includegraphics[width=1\textwidth]{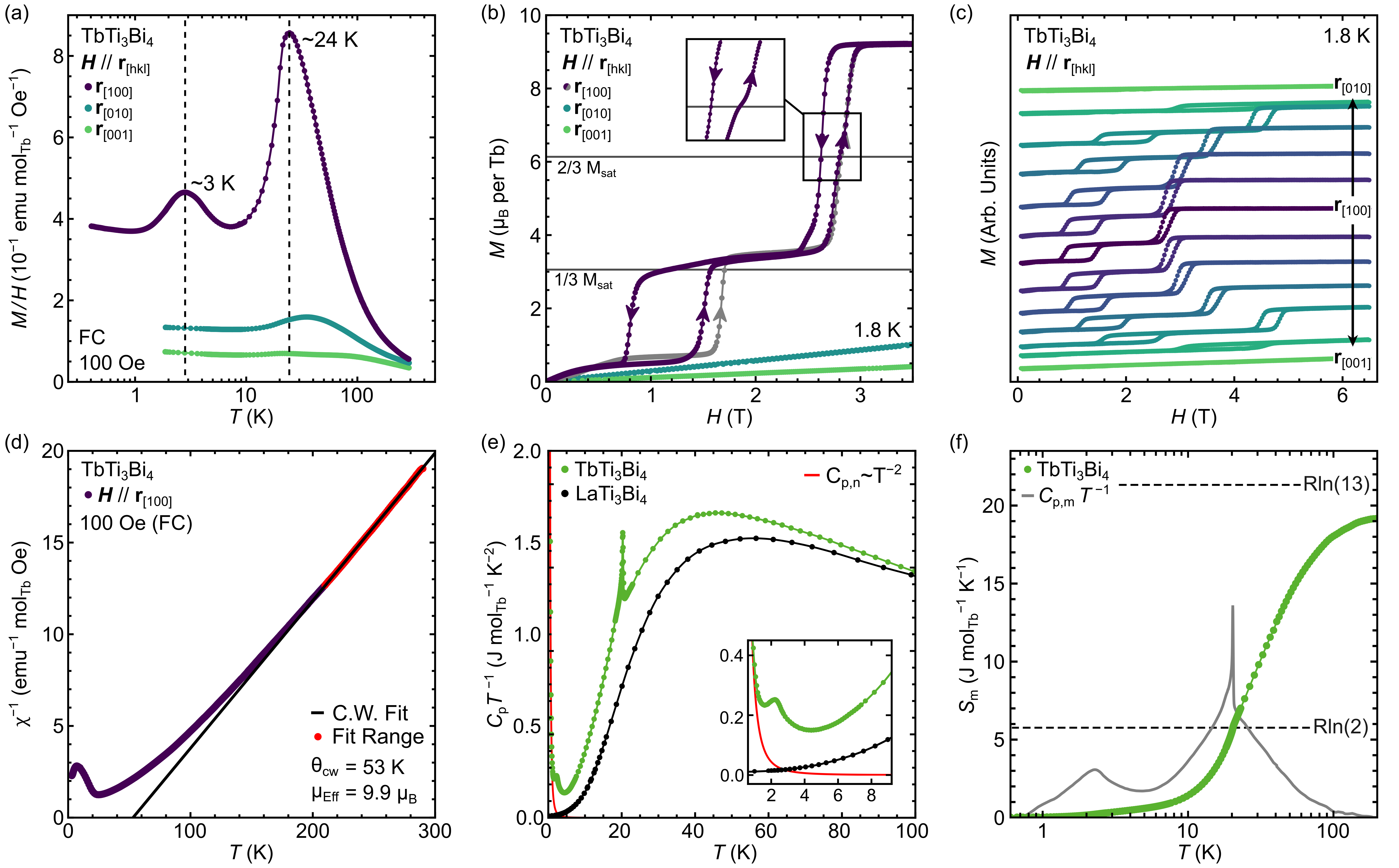}
\caption{(a) Magnetization data for single crystal TbTi$_3$Bi$_4$ with \textit{H} (100~Oe) oriented along the three primary crystallographic axes. We observe two magnetic transitions at 24~K and 3~K into an antiferromagnetic ground state. (b) Isothermal magnetization curves highlight hysteretic metamagnetic transitions when $H\parallel r_{[100]}$. The gray data indicates the zero-field-cooled (ZFC) virgin magnetization curve, with field ramp directions denoted by the arrows. Higher-resolution data (inset) highlights a brief plateau near 2/3 M$_\text{sat}$. (c) Isothermal magnetization curves collected at 1.8~K at the sample is rotated to align the three crystallographic axes with the magnetic field. For simplicity, no ZFC virgin curves are shown. (d) Inverse susceptibility (100~Oe) and corresponding Curie-Weiss analysis (black) on the highlighted temperature range (red) yields an effective moment consistent with Tb$^{3+}$. (e) Heat capacity and the resulting integrated entropy (f) yield results consistent with the Tb ion. LaTi$_3$Bi$_4$ was used as a nonmagnetic standard, and a simple $\sim$T$^{-2}$ model was used to remove the contribution from a nuclear Schottky.}
\label{fig:3}
\end{figure*}

Due to the reduced symmetry (\textit{Fmmm}), the \textit{Ln}Ti$_3$Bi$_4$ materials tend to be highly anisotropic, particularly with regards to the quasi-1D zig-zag chains. TbTi$_3$Bi$_4$ is no exception. To diagnose the potential anisotropies in the material, we first begin with a study of the magnetization along different principle crystallograpic axes. Figure \ref{fig:3}(a) presents the field-cooled (FC) magnetization curve of a single crystal of TbTi$_3$Bi$_4$ with the magnetic field (100~Oe) parallel to the [100], [010], and [001] directions. At these low fields, there is no difference between the field-cooled (FC) and zero-field-cooled (ZFC) measurements. TbTi$_3$Bi$_4$ appears to adopt an antiferromagnetic ground state, with two transitions at 24~K and 3~K. The [100] direction corresponds to the ``easy axis,'' and orienting the magnetic field parallel to the [010] or [001] induces a substantially weaker effect.

Figure \ref{fig:3}(b) demonstrates the isothermal magnetization of TbTi$_3$Bi$_4$ with the magnetic field parallel to the [100], [010], and [001] directions at 1.8~K. Due to the hysteretic nature of TbTi$_3$Bi$_4$, care must be taken to control the magnetic and thermal history of the samples. For example, Figure \ref{fig:3}(b) ($H\parallel r_{[100]}$) shows three different field paths: 1) ZFC virgin magnetization data (gray), 2) decreasing field (purple, down arrow), and 3) increasing field (purple, up arrow). The virgin curve is only obtained upon the first field sweep after cooling below 24~K with no applied field (ZFC). All subsequent field sweeps will fall onto the purple data, tracing either the field increasing (up arrow) or field decreasing (down arrow) path. Continuing the data into the negative field direction is not necessary as the system exhibits no coercivity at zero-field, and loops collected from 0~T to 7~T are functionally identical to those from 0~T to $-$7~T. For the vast majority of our analysis we will omit the virgin magnetization curves. The differences between the ZFC virgin curve and the next positive field sweep are likely due to the domain structure that forms on cooling in zero field, and are not the focus of this work.

The isothermal magnetization data saturates at approximately 9.2~$\mu_\text{B}$ by 3~T, in good agreement with the M$_\text{Sat.}=9~\mu_\text{B}$ expected for Tb$^{3+}$. The first metamagnetic transition plateaus at a moment of 2.9~$\mu_\text{B}$--3.2~$\mu_\text{B}$, which is approximately 1/3 M$_\text{Sat.}$. Closer inspection midway through the second metamagnetic transition reveals an unusual inflection (see Figure \ref{fig:3}(b,inset)). This inflection occurs at approximately 6.2~$\mu_\text{B}$, slightly above 2/3 M$_\text{Sat.}$. This inflection and other properties of the isothermal magnetization curves will be studied in the following figure. As implied by the temperature-dependent magnetization shown in Figure \ref{fig:3}(a), the magnetic response is suppressed quickly as we rotate the field away from the [100] direction. Figure \ref{fig:3}(c) provides a series of magnetization curves collected while rotating between $H\parallel r_{[010]}$, $r_{[100]}$, and $r_{[001]}$ in 15\degree~ increments. Only the increasing and decreasing field directions are shown (no ZFC virgin magnetization curve). The angular dependence is relatively simple, with the transitions being stretched towards higher fields with increasing deviations from $H\parallel r_{[100]}$, again suggesting strong easy-plane (and further, easy-[100]) anisotropy. 

Figure \ref{fig:3}(d) presents the inverse susceptibility, collected under a 100~Oe (FC) field where $H\parallel r_{[100]}$. The high temperature data exhibits a slight upwards curvature, potentially due to a temperature-independent contribution from core diamagnetism (bismuth) or the quartz sample holder. To account for the curvature, we attempted fits utilizing a temperature-independent $\chi_0$ term, though the result was an nonphysical effective moment of $\sim$11~$\mu_\text{B}$. We opt instead to utilize a limited high-temperature regime (red) for a simpler Curie-Weiss fit (black), which yields a $\theta_\text{cw}$ =53~K and a $\mu_\text{Eff}$ = 9.9$\mu_\text{B}$. The effective moment agrees well with that expected of Tb$^{3+}$ (9.72$\mu_\text{B}$). However, the Weiss temperature is unusually large (53~K), likely affected by the temperature independent contribution, limited temperature range of the fit, and crystal-field effects.

\begin{figure*}
\includegraphics[width=1\textwidth]{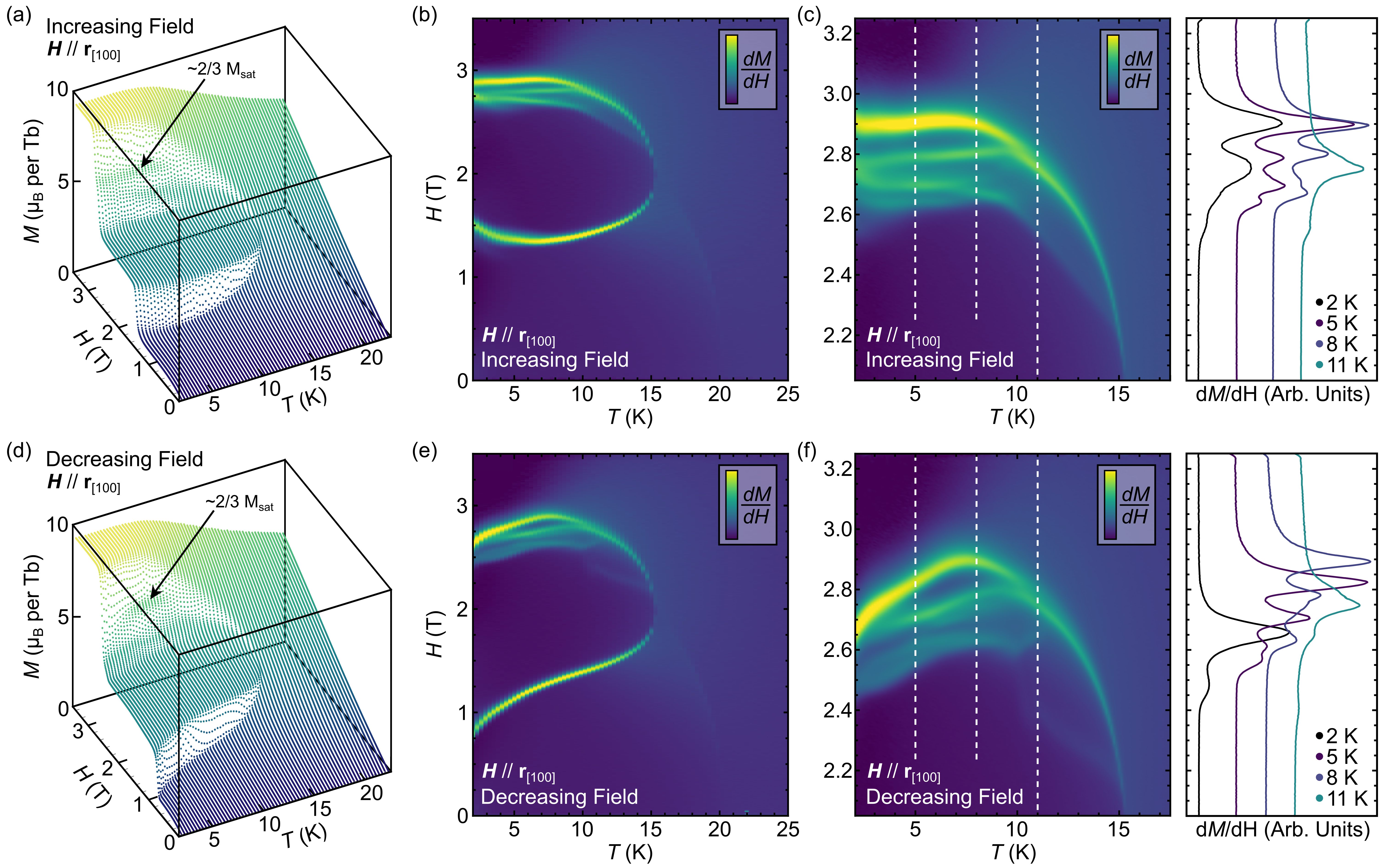}
\caption{(a,d) 3D perspective plot for isothermal magnetization curves ($H\parallel r_{[100]}$) for TbTi$_3$Bi$_4$ collected for both the increasing (a) and decreasing (d) field directions. Many subtle changes in the magnetization occur in proximity to the plateau at 2/3 M$_\text{Sat.}$, particularly in the intermediate temperature regime (2~K--15~K). (b,e) Phase pocket boundaries in the temperature-field phase diagrams are highlighted by taking the field derivative of the isothermal magnetization data. (c,f) Higher resolution temperature-field phase diagram data collected to exemplify the sheer complexity and number of unique states in TbTi$_3$Bi$_4$ near 2/3 M$_\text{Sat.}$. Several line cuts through the data (white dashed) are shown to the right to help delineate between phase boundaries.}
\label{fig:4}
\end{figure*}

Figure \ref{fig:3}(e) presents the zero-field heat capacity (plotted in $C_\text{p}T^{-1}$) data collected on a single crystal of TbTi$_3$Bi$_4$ down to 100~mK with measurements of LaTi$_3$Bi$_4$ used as a nonmagnetic standard. Two data sets were combined to construct this data set, a conventional helium measurement (200~K -- 1.8~K) and a measurement performed with a dilution refrigerator insert (4~K -- 100~mK). The data sets did not require substantial adjustment or scaling to overlap in the region from 4~K -- 1.8~K after normalization by the sample mass. The two magnetic transitions are evident around 20~K and 2~K. At temperatures below 2~K, a nuclear Schottky peak dominates the heat capacity signal. The peak is still increasing at 0.2~K, which agrees relatively well with a Tb-based Schottky peak ($T_\text{Sch.,Tb}\sim0.1$~K).\cite{heltemes1961nuclear} Utilizing LaTi$_3$Bi$_4$ as the nonmagnetic standard (black) and invoking a simple nuclear Schottky model $C_\text{p,n}\sim T^{-2}$, we can extract the magnetic entropy for TbTi$_3$Bi$_4$. 

Figure \ref{fig:3}(f) shows the integrated magnetic entropy (green) as a function of temperature. The extracted magnetic contribution to the heat capacity ($C_\text{p,m}T^{-1}$) is also shown (gray) as a schematic reference for the magnetic transition temperatures (not to scale). Interestingly, integrating $C_\text{p,m}$ up to the primary magnetic transition appears to correspond with an entropy release of approximately $R\ln 2$. Contributions to the magnetic entropy gradually continue up to 100~K, however, consistent with observations in other \textit{Ln}Ti$_3$Bi$_4$ materials.\cite{ortiz2023evolution} By 100~K we recover approximately 90\% of the total $R\ln 13$ entropy expected for Tb$^{3+}$. Considering that the higher temperature heat capacity data tends to be noisier (and that the nonmagnetic subtraction is more prone to error at high temperatures), this is remarkably good agreement with expectations.

We now return to the hysteretic metamagnetic transitions observed in Figure \ref{fig:3}(b) and the brief plateau near 2/3 M$_\text{Sat.}$ (Figure \ref{fig:3}(b,inset)). Closer inspection of the second metamagnetic transition (H$\sim$ 2.8--3.0~T) revealed a much more complex magnetic landscape than the 1.8~K isothermal curve would suggest. Figure \ref{fig:4}(a,d) present our efforts to map the magnetic landscape of TbTi$_3$Bi$_4$, demonstrating a set of isothermal magnetization curves collected in both the increasing (a) and decreasing (d) field directions. As mentioned previously, care must be taken to account for the thermal and magnetic history of TbTi$_3$Bi$_4$ samples. Each isothermal curve was performed by warming the sample above 50~K, and then cooling to the desired temperature under zero-field conditions (ZFC). As such, the magnetic history is erased for each measurement. No ZFC virgin magnetization curves are shown here, but they have been included in the ESI for completeness.

The sharpness of the second transition belies a network of subtle changes in the magnetization. These transitions are quantitatively consistent between samples and between sample batches, suggesting that the phenomenon observed in Figure \ref{fig:3} and Figure \ref{fig:4} are intrinsic to TbTi$_3$Bi$_4$. Taking the field-derivative of the magnetization data shown in (a,d) yields the phase diagrams shown in Figure \ref{fig:4}(b,e). Even at this scale, the plethora of phase pockets near 2/3 M$_\text{Sat.}$ is immediately evident, though they are most pronounced at intermediate temperatures from 2~K--15~K. 

To better highlight the intricacy of the magnetic landscape near the 2/3 M$_\text{Sat.}$ plateau, we collected higher density data (Figure \ref{fig:4}(c,f)) over limited temperature [2~K--18~K] and field [2~T--3.5~T] ranges. The magnetic history was erased between each isothermal curve as described previously. ZFC virgin curves have been omitted, though they have been included in the ESI. While the number of pockets appears to be roughly conserved, the qualitative behavior of the diagram depends strongly on whether the data is collected upon increasing or decreasing field. A series of line cuts at various temperatures (white dashed lines) have been included alongside both phase diagrams to help visualize the phase boundaries. Considering the complexity of the magnetic landscape, neutron diffraction experiments will be invaluable to identify the types and natures of the individual magnetic phases.

\begin{figure*}
\includegraphics[width=1\textwidth]{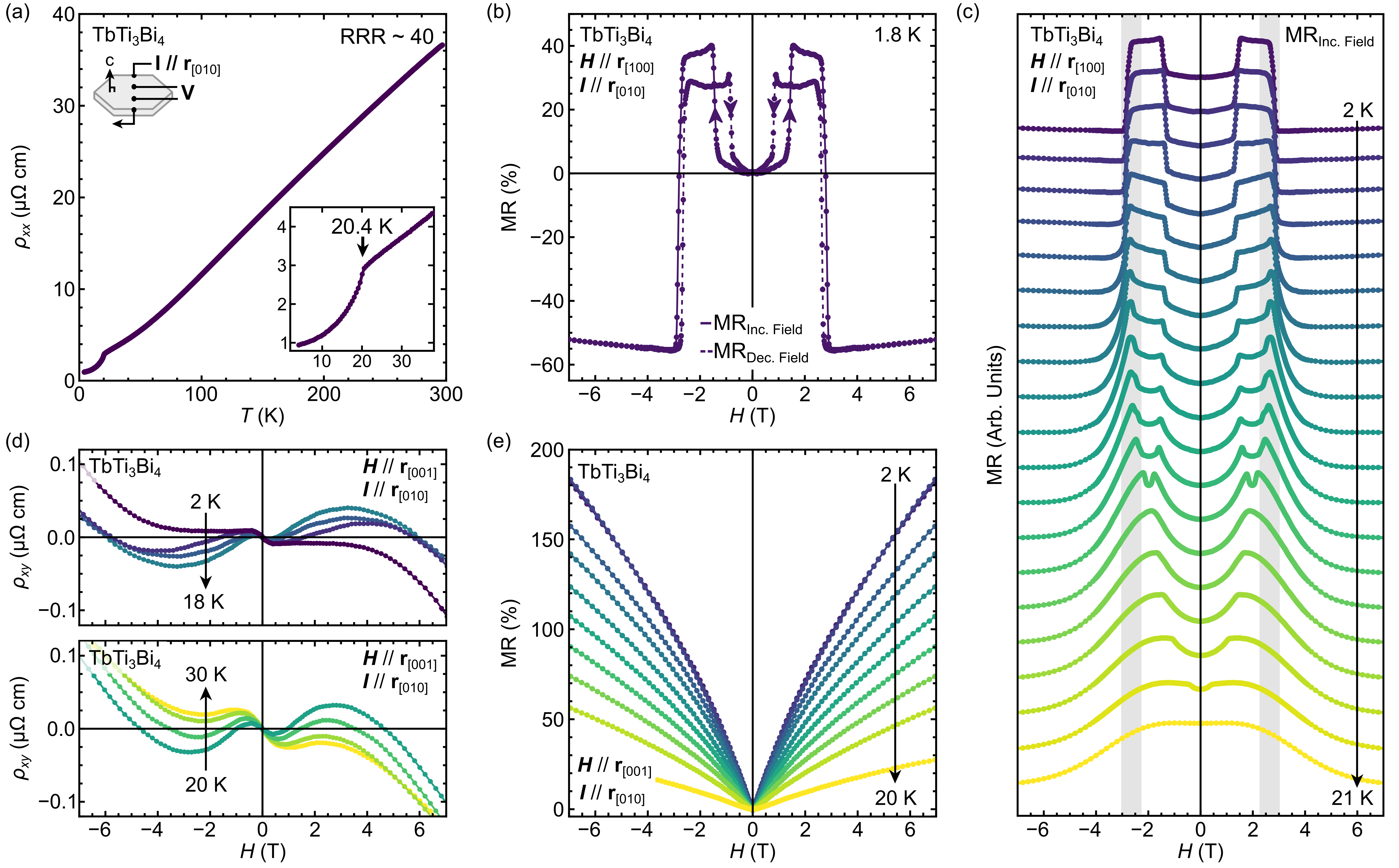}
\caption{(a) Zero-field electrical resistivity measurements on TbTi$_3$Bi$_4$ with $I\parallel r_{[010]}$. The sample quality is good (RRR $\sim$ 40), and the magnetic transition is clearly observed at 20.4~K. (b) Magnetoresistance (MR) measurements performed at 1.8~K with $H\parallel r_{[100]}$ and $I\parallel r_{[010]}$. Clear transitions in the MR are observed, coinciding with the metamagnetic transitions. Upon reaching the second transition, the system switches from a substantial positive (+40\%) MR to a strong negative MR ($-$50\%). (c) The temperature evolution of the transverse MR exhibits subtle changes as we approach the plateau at 2/3 M$_\text{Sat.}$. The gray regions highlight the corresponding field ranges where the multiple phase pockets are observed in magnetization measurements. (d) Hall effect measurements with $H\parallel r_{[001]}$ and $I\parallel r_{[010]}$. Strong non linearity is observed, but is currently ascribed to multiband effects -- particularly since no change is observed moving through the Neel temperature. (e) Strong positive MR is observed for samples with $H\parallel r_{[001]}$ and $I\parallel r_{[010]}$, in stark contrast to the negative MR observed in (b).
}
\label{fig:5}
\end{figure*}

Considering the complexity of the magnetic landscape, we were curious whether the features observed in Figure \ref{fig:3} and Figure \ref{fig:4} would have a dramatic influence on the magnetotransport. Cleavable antiferromagnetic materials have recently been a focus in fields like spintronics,\cite{jungwirth2016antiferromagnetic,baltz2018antiferromagnetic} and while the low-temperature magnetism of the rare-earth metals may prevent immediate practical application of these materials, there is still insight to be gleaned from the transport data. In particular, probing the connection between complex magnetic phases (\textit{via} the supporting sublattice) and the transport properties (dominated by the kagome Ti states at E$_\text{F}$) provides insight into how the sublattices interact. 

Figure \ref{fig:5}(a) shows the zero-field, in-plane resistivity of TbTi$_3$Bi$_4$ with $I\parallel r_{[010]}$ (8~mA). A simplified schematic of our sample setup is shown in the upper corner of Figure \ref{fig:5}(a). While the schematic draws the hexagonal faceted crystal to highlight the [010] facet (edge), the actual sample employed in the measurements was a bar cut from the bulk crystal with a length of 0.4~mm and a cross-sectional area of 0.05~mm$^2$. The crystal quality is good (RRR$\sim$40) and the zero-field transport is indicative of a relatively high mobility metal. The magnetic transition at 20~K is evident in the resistivity, highlighted by a sharp cusp and subsequent drop. This is perhaps the expected result for ordering from the paramagnetic regime, sharply reducing spin-scattering. 

We now apply a magnetic field ($H\parallel r_{[100]}$) to examine the in-plane (transverse) magnetoresistance (MR) of TbTi$_3$Bi$_4$. Figure \ref{fig:5}(b) shows the symmetrized transverse MR with $H\parallel r_{[100]}$ and $I\parallel r_{[010]}$ (8~mA) at 1.8~K. As before, we are primarily interested in the magnetotransport under repeated field cycles and have omitted the ZFC virgin curves. Instead, these measurements begin in the fully polarized (9~T) state,  ramping to $-$9~T, and finally back to 9~T. Symmetrized curves corresponding to increasing field direction (in both positive and negative directions) have been indicated by solid purple lines, whereas the decreasing field direction is represented by dashed purple lines. As one would expect from the magnetization results, we observe two sharp transitions in the MR, one for each of the metamagnetic features. The first metamagnetic transition marks a sharp increase in the MR of approximately 30--40\%. The second metamagnetic transition marks a sharp decrease in the MR to approximately $-$50\%, which one may expect for a total polarization of the Tb moments. We find the sharp switching between positive and negative MR intriguing in TbTi$_3$Bi$_4$, particularly for a system which is already quite metallic ($\sim$1$\mu$Ohm-cm at 1.8~K).

The strong MR response in TbTi$_3$Bi$_4$ at 1.8~K highlights the sensitivity of the transport to the particular magnetic state. Given the complexity of the magnetic phase pockets near the plateau at 2/3 M$_\text{Sat.}$ (see Figure \ref{fig:4}(c,f)), we were curious whether the electronic transport would be sensitive to the subtle shifts in the bulk magnetization. Figure \ref{fig:5}(c) provides a series of MR curves as a function of temperature. For simplicity and clarity, we are only showing the data under increasing fields, though full field loops are shown in the ESI. The field range corresponding to the 2/3 M$_\text{Sat.}$ plateau (see Figure \ref{fig:3}(b)) has been accentuated by the light gray background. Immediately before the transition to negative MR, we observe that the MR cusp becomes a peak -- then splits into two/three peaks -- merge back together -- and then vanish as the system approaches the N\'eel temperature. These effects occur at the intermediate temperatures consistent with the appearance of the various phase pockets in Figure \ref{fig:4}. This further suggests that these are bulk effects corresponding to distinct types of magnetic order. 

Utilizing the same sample geometry as Figure \ref{fig:5}(a--c), we turned to investigate the Hall effect measurement with $H\parallel r_{[001]}$ and $I\parallel r_{[010]}$. It is critical to note that there are no metamagnetic transitions with $H\parallel r_{[001]}$, so both the MR and Hall effect are probing a much simpler magnetic landscape. Furthermore, the effect of hysteresis (and the sensitivity to field and thermal history) vanishes. Figure \ref{fig:5}(d) shows the antisymmetrized Hall resistivity for TbTi$_3$Bi$_4$ over two temperature regimes. The top figure shows the Hall response within the antiferromagnetic state (2~K -- 18~K), and the bottom figure highlights temperatures within the paramagnetic phase (20~K -- 30~K).

As with the MR, full field loops were collected at each temperature, starting at the fully-polarized 7~T state, ramping to $-$7~T, and finally back to 7~T. No subtractions or corrections have been performed besides antisymmetrization. All data sets shown here exhibit strongly non-linear behavior. A signal that looks curiously like an anomalous Hall signal is evident at 2~K, though the signal persists in shape and magnitude above the magnetic transition temperature. Furthermore, at high temperatures we recover an additional inflection in the data. The signal is also relatively small, leading us to suspect that we may be seeing the effect of multiband transport. Finally, Figure \ref{fig:5}(e) presents the transverse MR when $H\parallel r_{[001]}$ and $I\parallel r_{[010]}$. In the absence of the metamagnetism, no hysteresis in the MR is observed. Further, the system exhibits only strong positive MR ($\sim$200\%) over all fields and temperatures.

Nearly simultaneously with this work, a manuscript focusing on the magnetotransport and ARPES in TbTi$_3$Bi$_4$ were posted.\cite{cheng2024giant} This work complements the results shown here by focusing on the Hall effect when $H\parallel r_{[100]}$ (inducing the metamagnetic transitions). The authors note an record-high anomalous Hall conductivity, nearly an order of magnitude larger than comparable systems. 

Overall, our work into TbTi$_3$Bi$_4$ has revealed a unique antiferromagnetic, cleavable metal with an exceedingly complex landscape of magnetic phases. The magnetic and transport properties are rich and diverse, with multiple unique magnetic phases to investigate, strong switching from positive to negative MR near the 2/3 M$_\text{Sat.}$ plateau, and a complex coupling between the magnetic phase pockets and the transport. We now turn to briefly investigate the magnetic properties of the \textit{Ln}$_{2-x}$Ti$_{6+x}$Bi$_9$ shurikagome metals and the influence of disorder on the \textit{Ln}--\textit{Ln} chains.

\subsection{Properties of \textit{Ln}$_{2-x}$Ti$_{6+x}$Bi$_9$ (\textit{Ln}: Tb--Lu) family} 

When we first identified the \textit{Ln}$_{2-x}$Ti$_{6+x}$Bi$_9$ shurikagome metals, we were surprised by the substitution of \textit{Ln} into the Ti$_8$Bi$_9$ structure. Unlike integration of the rare-earth elements into frameworks like CoSn, there are no clear interstitial voids in Ti$_8$Bi$_9$. Further, the miscibility of most rare-earth elements into Ti is low, and it is initially unintuitive that \textit{Ln}$_\text{Ti}$ is the predominant defect. The incomplete substitution of \textit{Ln} is also perplexing. 

While single crystal X-ray diffraction (SCXRD) results can be solved using fully occupied \textit{Ln} sites (forming fully occupied chains) with relatively favorable statistics, large amounts of excess electron density accumulate on the \textit{Ln}4 and \textit{Ln}5 sites. The residual Q-peaks can be eliminated by allowing the occupancy of the \textit{Ln} sites to vary, which results in partially occupied \textit{Ln} chains. This is also consistent with energy dispersive spectroscopy (EDS) results, which support a \textit{Ln} content from 6--9\% (ideal \textit{Ln}$_{2}$Ti$_{6}$Bi$_9$ is 12\% \textit{Ln}). For simplicity, we have restrained the chain sites to remain fully occupied (e.g. Occ$_\text{\textit{Ln}4,Ti}$ + Occ$_\text{\textit{Ln}4,\textit{Ln}}$ = 1). Even this appears to slightly underestimate the amount of Ti, and EDS results suggest that some Ti may be substituting elsewhere as well. The published Ti--Bi phase diagram supports Ti$_\text{Bi}$ substitution, as Ti$_8$Bi$_9$ supports a full solid solution from 50:50 Ti:Bi to 35:65 Ti:Bi.\cite{vassilev2006contribution} Regardless, the SCXRD data did not improve substantially with more complex substitution models. SCXRD tables, CIF files, and EDS results for the entire series can be found in the ESI.

The substitution of the \textit{Ln} atoms creates a clear separation between the \textit{Ln}4/\textit{Ln}5 sites and the Ti-based shurikagome network. As suggested by our designation ``shurikagome,'' the coordination of the Ti sublattice in the \textit{Ln}$_{2-x}$Ti$_{6+x}$Bi$_9$ compounds contains pieces of the corner-sharing triangular motif observed in prototypical kagome samples. The parent structure Ti$_8$Bi$_9$ has been known for some time,\cite{richter1997preparation} though the electronic structure has only been calculated through high-throughput compendiums like Materials Project.\cite{Jain2013} In an effort to understand the underlying electronic structure of the \textit{Ln}$_{2-x}$Ti$_{6+x}$Bi$_9$ compounds, we have performed first-principle density functional theory calculations on the ideal \textit{Ln}$_{2}$Ti$_{6}$Bi$_9$ structure. To avoid the additional complication of magnetism, we examine the electronic structure of Lu$_2$Ti$_6$Bi$_9$ as a proxy for the material family.

\begin{figure}
\includegraphics[width=\linewidth]{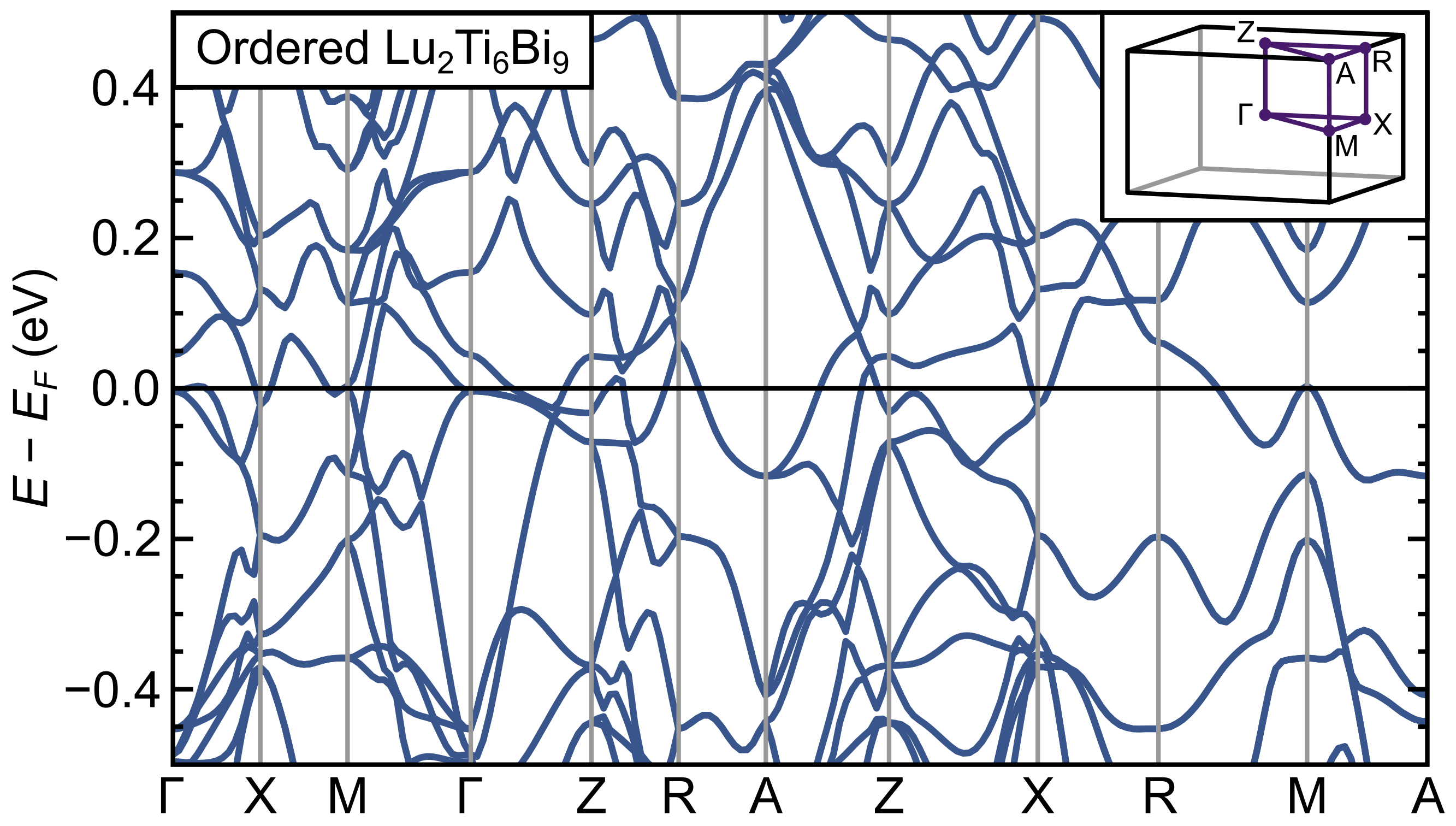}
\caption{As a proxy for the \textit{Ln}$_{2-x}$Ti$_{6+x}$Bi$_9$ compounds, we have computed the electronic structure for the idealized, ordered compound Lu$_2$Ti$_6$Bi$_9$. There are several interesting features near the Fermi level, including a Fermi-level band edge and associated ``flat band'' centered at $\Gamma$ and extending towards the $X$ and $Z$ high symmetry points.}
\label{fig:ele}
\end{figure}

\begin{figure*}
\includegraphics[width=1\textwidth]{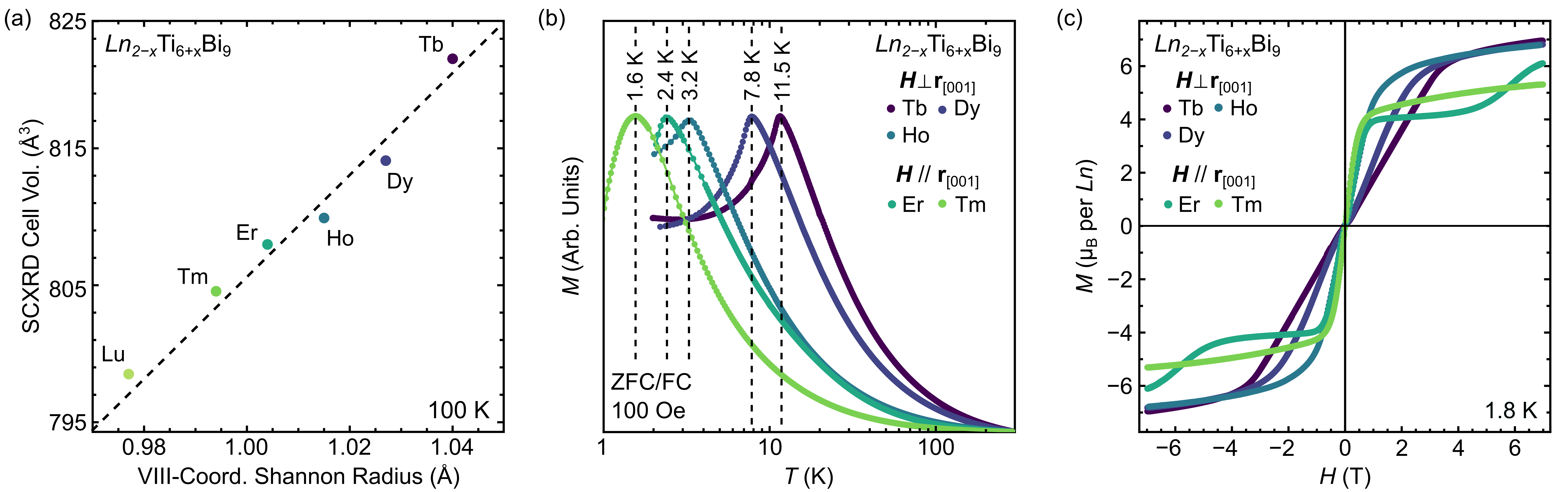}
\caption{(a) Summary of the single crystal diffraction data, showing a linear trend between the SCXRD cell volume and the 8-coordinate Shannon radius. (b) Overview of the temperature-dependent magnetization in crystals of \textit{Ln}$_{2-x}$Ti$_{6+x}$Bi$_9$. A cusp-like feature emerges for samples of Tb--Er with $H\perp r_{[100]}$ while Er--Tm exhibit anomalies with $H\parallel r_{[100]}$. The agglomerated data is shown in arbitrary units to aid in comparison of the transition temperature and the qualitative behavior. Both zero-field-cooled (ZFC) and field-cooled (FC) measurements are shown for samples with Tb--Er, and none exhibit ZFC/FC splitting. (c) Isothermal magnetization curves for \textit{Ln}$_{2-x}$Ti$_{6+x}$Bi$_9$ samples along the same direction as in the temperature-dependent magnetization. The data is normalized based on the ideal formula unit \textit{Ln}$_{2}$Ti$_{6}$Bi$_9$, which underestimates the saturation magnetization, but can be compared to the expected results in Table \ref{tab:1}. The Lu-based compound is nonmagnetic and has been omitted from these datasets.}
\label{fig:6}
\end{figure*}

Figure \ref{fig:ele} provides the calculated electronic structure for Lu$_2$Ti$_6$Bi$_9$ alongside a schematic diagram of the \textit{P}4\textit{/nmm} Brilluoin zone.\cite{setyawan2010high} There are several interesting features apparent within the data. We first note a Fermi-level band edge and associated ``flat--band,'' in two distinct directions, centered at $\Gamma$ ($\Gamma-X, \Gamma-Z$). The extent of this flat band is relatively limited when compared to the prototypical kagome band structure, though the proximity to the Fermi level is interesting. Similar to the case of the \textit{Ln}Ti$_3$Bi$_4$ compounds, the states near the Fermi level are almost entirely comprised of Ti-based orbitals arising from shurikagome motif. Beyond the flat-bands, the Lu$_2$Ti$_6$Bi$_9$ band structure contains many dispersive Dirac-like features and several avoided crossings near E$_\text{F}$ (at and near the $M$-point, $\sim$10~meV above E$_\text{F}$ between $Z$ and $R$, $\sim$50~meV above E$_\text{F}$ between $A$ and $Z$). 

Flat bands and/or van Hove singularities near E$_\text{F}$ are commonly cited as the cause for a wide array of unusual behavior, presumably due to the localized nature of electrons in flat-bands and associated correlative effects. The numerous band edges at or very near ($\pm$25~meV) the Fermi level in Lu$_2$Ti$_6$Bi$_9$ are suggestive of numerous Fermi surface Lifshitz transitions, which supports the off-stoichiometric behavior of these compounds. This additionally indicates a complex, highly sample-dependent Fermi-surface structure. Further, while the spectral weight associated with the Fermi-level ``flat-bands'' in Lu$_2$Ti$_6$Bi$_9$ is relatively small, it is still suggestive of potential novel behavior potentially observable via ARPES. 

Out of curiosity, we performed preliminary ARPES experiments on an in-house laser ARPES system on crystals of Tb$_{2}$Ti$_{6}$Bi$_9$ (which has the highest rare-earth occupancy) at approximately 10~K. We were able to observe a high density of states near the Fermi level, consistent with Figure \ref{fig:ele} and the associated ``flat-bands.'' However, the intrinsic disorder in the \textit{Ln}$_{2-x}$Ti$_{6+x}$Bi$_9$ compounds is currently prohibitive to performing in-depth ARPES investigations, particularly considering the relatively complex electronic structure. Whether the flat bands can be directly ascribed to the Ti-based ``shurikagome'' sublattice (in analogy to tight binding calculations in the prototypical kagome metals) is a subject of current investigation. We believe that the applicability of both the DFT and ARPES models will require parallel efforts to synthesize compounds with reduced disorder. The existence of the halfnium-based Hf$_8$Bi$_9$\cite{richter1997preparation} and some similar vanadium-based V$_8$Sb$_9$\cite{kleinke1998stabilization} compounds may offer some flexibility in the choice of framework cations that may offer routes to achieve fully-occupied \textit{Ln}$_{2}$M$_{6}$X$_9$ samples.

Indeed, synthesizing fully-occupied Ti$_8$Bi$_9$-based shurikagome compounds may be challenging, as the extent of disorder in samples of \textit{Ln}$_{2-x}$Ti$_{6+x}$Bi$_9$ appear to vary quite strongly with the rare-earth species. Figure \ref{fig:6}(a) highlights the cell volume for the \textit{Ln}$_{2-x}$Ti$_{6+x}$Bi$_9$ shurikagome metals as a function of the 8-coordinate Shannon radius. Despite the variable rare-earth occupancy, the compounds still adhere to a nearly linear expansion in cell volume, as one would expect of a rare-earth series. These results are briefly summarized in Table \ref{tab:1}. The columns labeled Occ$^\text{scxrd}_\text{\textit{Ln}4}$ and Occ$^\text{scxrd}_\text{\textit{Ln}5}$ indicate the refined occupancy of the chain sites \textit{Ln}4 and \textit{Ln}5. The sum of the SCXRD-determined occupancies is provided in the column labeled ``Tot \textit{Ln}$^\text{scxrd}$.'' This can be directly compared with the composition extracted from EDS measurements in the next column.

Some interesting trends appear in the agglomerate data. First, SCXRD appears to consistently calculate higher concentrations of \textit{Ln} when compared to EDS. However, regardless of technique, both the \textit{Ln4} and \textit{Ln5} site occupancies decrease as we move to smaller \textit{Ln} atoms. This effect is substantially more pronounced on the \textit{Ln5} (smaller) rare-earth site, where the occupancy is nearly halved by the time we reach Lu$_{2-x}$Ti$_{6+x}$Bi$_9$. We performed some exploratory syntheses around Tb$_{2-x}$Ti$_{6+x}$Bi$_9$ in an attempt to fully saturate the \textit{Ln} site, though only small changes in the rare-earth composition were noted. Coincidentally, we have not observed the formation of \textit{Ln}$_{2-x}$Ti$_{6+x}$Bi$_9$ with large rare-earths (La--Gd). We speculate that the larger rare-earths may destabilize the structure. However, under this assumption, we find it unusual that smaller rare-earth elements have decreasing occupancies, particularly on the smaller site. 

Remarkably, despite the high degree of chemical disorder in crystals of \textit{Ln}$_{2-x}$Ti$_{6+x}$Bi$_9$, magnetization measurements exhibit properties more indicative of antiferromagnetic order. Due to the tetragonal (\textit{P}4/\textit{nmm}) structure and the quasi-1D \textit{Ln}--\textit{Ln} chain, there are two directions of interest for magnetization measurements, $H\parallel r_{[001]}$ (parallel to chain) and $H\perp r_{[001]}$ (perpendicular to chain). Samples containing Tb, Dy, and Ho exhibit a cusp in the magnetization when $H\perp r_{[001]}$, whereas Er and Tm have features when $H\parallel r_{[001]}$ instead. All compounds exhibit weak anisotropy with the exception of Tm. When $H\parallel r_{[001]}$,  Tm$_{2-x}$Ti$_{6+x}$Bi$_9$ exhibits an order of magnitude stronger response. These results are summarized in Figure \ref{fig:6}(b) which provides an overview of the magnetization for the \textit{Ln}$_{2-x}$Ti$_{6+x}$Bi$_9$ shurikagome metals along their ``favored'' direction. Note that Figure \ref{fig:6}(b) is shown in arbitrary units for ease of comparison, though full data sets are provided in the ESI.

\setlength{\tabcolsep}{5.5pt} 
\renewcommand{\arraystretch}{1} 
\begin{table*}[]
\begin{tabular}{@{}lcccccllcclll@{}}
\toprule
Shurikagome \textit{Ln}  & Occ$^\text{scxrd}_\text{\textit{Ln}4}$ & Occ$^\text{scxrd}_\text{\textit{Ln}5}$   & Tot \textit{Ln}$^\text{scxrd}$    & Tot \textit{Ln}$^\text{eds}$  & M$^\text{scxrd}_\text{Sat}$ ($\mu_\text{B}$) & M$^\text{eds}_\text{Sat}$ ($\mu_\text{B}$) & M$_\text{Sat}$ ($\mu_\text{B}$)  & T$_\text{N}$ (K) & $\mu_\text{cw}$ ($\mu_\text{B}$)& $\theta_\text{cw}$ (K)\\ \midrule
Tb$_{2-x}$Ti$_{6+x}$Bi$_9$    & 0.89 & 0.76 & 1.65 & 1.4 & 7.4     &6.3      & 7.0   & 11.5   & 9.8 &  21      \\
Dy$_{2-x}$Ti$_{6+x}$Bi$_9$    & 0.84 & 0.62 & 1.46 & 1.3 & 6.8   & 5.9     & 6.8  & 7.8     & 9.6 & 14       \\
Ho$_{2-x}$Ti$_{6+x}$Bi$_9$    & 0.83 & 0.56 & 1.39 & 1.2 & 6.9   & 6.0     & 6.9  & 3.2     & 9.9 & 0.8      \\
Er$_{2-x}$Ti$_{6+x}$Bi$_9$    & 0.81 & 0.54 & 1.35 & 1.2 & 6.1   & 5.3     & 6.2  & 2.4     & 10.5 & 1.0       \\
Tm$_{2-x}$Ti$_{6+x}$Bi$_9$   & 0.81 & 0.56 & 1.37 & 1.2 & 4.8   & 4.3     & 5.3  & 1.6     & 8.6 & 1.1       \\  
Lu$_{2-x}$Ti$_{6+x}$Bi$_9$   & 0.75 & 0.46 & 1.21 & 1.0 & N/A   & N/A     & N/A  & N/A     & N/A & N/A      \\ \bottomrule\\
\end{tabular}
\caption{\label{tab:1} Summary of select results for the \textit{Ln}$_{2-x}$Ti$_{6+x}$Bi$_9$ ``shurikagome'' metals compounds investigated in this study. Of particular interest is the variable occupancy on the two rare-earth chain sites \textit{Ln4} and \textit{Ln5} as derived from single-crystal X-ray diffraction (SCXRD). We also provide comparisons between the nominal \textit{Ln} content as derived from both SCXRD and energy dispersive spectroscopy (EDS), and the expected saturation magnetization based on the respective nominal compositions.}
\end{table*}

The data presented in Figure \ref{fig:6}(b) includes both ZFC and FC data for Tb--Er samples. The lack of ZFC/FC splitting is surprising, as we may reasonably expect the strong disorder to result in a glassy magnetic state. Furthermore, AC susceptibility measurements on Tb$_{2-x}$Ti$_{6+x}$Bi$_9$ indicate no frequency dependence. However, this is not universal throughout the series, as Tm$_{2-x}$Ti$_{6+x}$Bi$_9$ exhibits some frequency dependence. Notably, however, Tm also possesses one of the lowest \textit{Ln} occupancies (Tm$_{1.4}$Ti$_{2.6}$Bi$_9$) across the series. The magnetization anomalies follow the temperature scale (de Gennes scaling) expected of the latter half of the rare-earth row, with Tb possessing the highest ordering temperature, and Tm possessing the lowest. The results are generally reminiscent of antiferromagnetic-like order, particularly considering the lack of ZFC/FC splitting. As an additional diagnostic tool, we performed zero-field heat capacity measurements (see ESI) on both Tb$_{2-x}$Ti$_{6+x}$Bi$_9$ and Er$_{2-x}$Ti$_{6+x}$Bi$_9$ crystals from 300~K to 1.8~K. Both samples show clear, albeit small, heat capacity anomalies. These results are in good agreement with magnetization results and further support some degree of non-glassy magnetic order in the series. 

Figure \ref{fig:6}(c) demonstrates the isothermal magnetization for all magnetic \textit{Ln}$_{2-x}$Ti$_{6+x}$Bi$_9$ compounds at 1.8~K. For simplicity we have opted to show the same field directions as in Figure \ref{fig:6}(b), though both the $H\parallel r_{[001]}$ and $H\perp r_{[001]}$ directions are included in the ESI. All compounds near saturation by 7~T, though Er$_{2-x}$Ti$_{6+x}$Bi$_9$ has not saturated, owing to an unusual inflection in the data near 4~T. Regardless, we can extract the approximate saturation magnetization for all compounds, which has been summarized in Table \ref{tab:1} (column M$_\text{Sat}$). Our quantitative analysis assumes the molar susceptibility can be expressed in terms of the ideal formula unit \textit{Ln}$_{2}$Ti$_{6}$Bi$_9$. This underestimates the true saturation magnetization, though it can be compared directly with the ``expected saturation magnetization from XRD and EDS results (M$^\text{xrd}_\text{Sat}$ and M$^\text{eds}_\text{Sat}$). Some other parameters of interest (e.g. $\mu_\text{Eff}$ and $\theta_\text{cw}$) have also been added to Table \ref{tab:1}. Of course, due to the disorder and variable number of \textit{Ln} sites, we consider these values as first approximations. 

Remarkably, despite the high levels of disorder, the majority of \textit{Ln}$_{2-x}$Ti$_{6+x}$Bi$_9$ appear to show magnetic properties consistent with non-glassy magnetic order. The lack of ZFC/FC splitting, the lack of frequency dependence in most samples, and the presence of well-defined heat capacity anomalies suggest some degree of conventional magnetic ordering. The disorder assuredly plays a non-trivial role in the manifestation of this order, and we cannot completely exclude the possibility of some glassiness. More detailed investigations into the magnetism and chemical disorder of the \textit{Ln}$_{2-x}$Ti$_{6+x}$Bi$_9$ family are clearly needed. Future work should focus on controlling chemical disorder in the shurikagome metals, particularly with an emphasis on realizing an idealized \textit{Ln}$_{2}$M$_{6}$X$_9$ analog.

\section{Conclusions}
We began this work by describing a modified synthesis procedure that is capable of subverting phase competition to realize TbTi$_3$Bi$_4$, the smallest rare-earth containing member of the \textit{Ln}Ti$_3$Bi$_4$ kagome metals. We described the rationale for the formation of TbTi$_3$Bi$_4$, including the increased competition with the shurikagome metals \textit{Ln}$_{2-x}$Ti$_{6+x}$Bi$_9$ (\textit{Ln}: Tb--Lu). Our work into both classes of materials have demonstrated that TbTi$_3$Bi$_4$ is an exceedingly interesting candidate material -- an exfoliable, antiferromagnetic kagome metal with a complex landscape of magnetic phases, metamagnetic transitions, and 1/3 and 2/3 magnetization plateaus. Transport measurements highlight the impact the rare-earth magnetism has on the kagome-dominated transport. Upon entering the magnetization plateau, the system sharply switches between positive (+40\%) and negative ($-$50\%) magnetoresistance in a system that is already quite metallic ($\sim$1~$\mu$Ohm~cm). We have also briefly examined the properties of the \textit{Ln}$_{2-x}$Ti$_{6+x}$Bi$_9$ shurikagome metals. Despite considerable disorder on the rare-earth sublattices, the systems exhibit magnetization responses reminiscent of non-glassy antiferromagnetic order. Potential routes towards fully-occupied \textit{Ln}$_{2}$M$_{6}$X$_9$ analogs may provide a cleavable system with an interesting electronic structure. Together, our results not only provide insight into a new kagome lattice and an unusual family of chemically-disordered magnets, but also illuminate a synthetic strategy to help realize new members of the \textit{A}\textit{M}$_3$\textit{X}$_4$ family.

\section{Acknowledgements}

This work was supported by the U.S. Department of Energy (DOE), Office of Science, Basic Energy Sciences (BES), Materials Sciences and Engineering Division. We thank Jong Keum and the X-ray laboratory of the Oak Ridge National Laboratory Spallation Neutron Source (SNS) for use of their Photonic Science Laue camera. Research directed by G.D.S. is additionally sponsored by the Laboratory Directed Research and Development Program of Oak Ridge National Laboratory, managed by UT-Battelle, LLC, for the US Department of Energy. This research used resources of the Compute and Data Environment for Science (CADES) at the Oak Ridge National Laboratory, which is supported by the Office of Science of the U.S. Department of Energy under Contract No. DE-AC05-00OR22725. Portions of this work were supported by the U.S. Department of Energy, Office of Science, National Quantum Information Science Research Centers, Quantum Science Center (ARPES, Q.L., R.G.M.)

\section{Supplementary Info.}
Supplementary crystallography tables and corresponding CIF files; Additional magnetic phase diagram utilizing zero-field-cooled virgin curves; Preliminary ARPES data on TbTi$_3$Bi$_4$; Additional magnetoresistance data with full temperature-field data set and full field loops; Table of shurikagome compositions measured by EDS; Preliminary ARPES data on Tb$_{2-x}$Ti$_{6+x}$Bi$_9$; Magnetization versus temperature data for \textit{Ln}$_{2-x}$Ti$_{6+x}$Bi$_9$ to scale; Isothermal magnetization data for \textit{Ln}$_{2-x}$Ti$_{6+x}$Bi$_9$; Curie-Weiss fits for \textit{Ln}$_{2-x}$Ti$_{6+x}$Bi$_9$; Select heat capacity results for Tb$_{2-x}$Ti$_{6+x}$Bi$_9$ and Tm$_{2-x}$Ti$_{6+x}$Bi$_9$

\bibliography{LnTi3Bi4}

\providecommand{\noopsort}[1]{}\providecommand{\singleletter}[1]{#1}%
\begin{thebibliography}{61}%
\makeatletter
\providecommand \@ifxundefined [1]{%
 \@ifx{#1\undefined}
}%
\providecommand \@ifnum [1]{%
 \ifnum #1\expandafter \@firstoftwo
 \else \expandafter \@secondoftwo
 \fi
}%
\providecommand \@ifx [1]{%
 \ifx #1\expandafter \@firstoftwo
 \else \expandafter \@secondoftwo
 \fi
}%
\providecommand \natexlab [1]{#1}%
\providecommand \enquote  [1]{``#1''}%
\providecommand \bibnamefont  [1]{#1}%
\providecommand \bibfnamefont [1]{#1}%
\providecommand \citenamefont [1]{#1}%
\providecommand \href@noop [0]{\@secondoftwo}%
\providecommand \href [0]{\begingroup \@sanitize@url \@href}%
\providecommand \@href[1]{\@@startlink{#1}\@@href}%
\providecommand \@@href[1]{\endgroup#1\@@endlink}%
\providecommand \@sanitize@url [0]{\catcode `\\12\catcode `\$12\catcode `\&12\catcode `\#12\catcode `\^12\catcode `\_12\catcode `\%12\relax}%
\providecommand \@@startlink[1]{}%
\providecommand \@@endlink[0]{}%
\providecommand \url  [0]{\begingroup\@sanitize@url \@url }%
\providecommand \@url [1]{\endgroup\@href {#1}{\urlprefix }}%
\providecommand \urlprefix  [0]{URL }%
\providecommand \Eprint [0]{\href }%
\providecommand \doibase [0]{https://doi.org/}%
\providecommand \selectlanguage [0]{\@gobble}%
\providecommand \bibinfo  [0]{\@secondoftwo}%
\providecommand \bibfield  [0]{\@secondoftwo}%
\providecommand \translation [1]{[#1]}%
\providecommand \BibitemOpen [0]{}%
\providecommand \bibitemStop [0]{}%
\providecommand \bibitemNoStop [0]{.\EOS\space}%
\providecommand \EOS [0]{\spacefactor3000\relax}%
\providecommand \BibitemShut  [1]{\csname bibitem#1\endcsname}%
\let\auto@bib@innerbib\@empty
\bibitem [{\citenamefont {Park}\ \emph {et~al.}(2021)\citenamefont {Park}, \citenamefont {Ye},\ and\ \citenamefont {Balents}}]{park2021electronic}%
  \BibitemOpen
  \bibfield  {author} {\bibinfo {author} {\bibfnamefont {T.}~\bibnamefont {Park}}, \bibinfo {author} {\bibfnamefont {M.}~\bibnamefont {Ye}},\ and\ \bibinfo {author} {\bibfnamefont {L.}~\bibnamefont {Balents}},\ }\bibfield  {title} {\bibinfo {title} {{Electronic instabilities of kagome metals: saddle points and Landau theory}},\ }\href@noop {} {\bibfield  {journal} {\bibinfo  {journal} {Phys. Rev. B}\ }\textbf {\bibinfo {volume} {104}},\ \bibinfo {pages} {035142} (\bibinfo {year} {2021})}\BibitemShut {NoStop}%
\bibitem [{\citenamefont {Wang}\ \emph {et~al.}(2013)\citenamefont {Wang}, \citenamefont {Li}, \citenamefont {Xiang},\ and\ \citenamefont {Wang}}]{PhysRevB.87.115135}%
  \BibitemOpen
  \bibfield  {author} {\bibinfo {author} {\bibfnamefont {W.-S.}\ \bibnamefont {Wang}}, \bibinfo {author} {\bibfnamefont {Z.-Z.}\ \bibnamefont {Li}}, \bibinfo {author} {\bibfnamefont {Y.-Y.}\ \bibnamefont {Xiang}},\ and\ \bibinfo {author} {\bibfnamefont {Q.-H.}\ \bibnamefont {Wang}},\ }\bibfield  {title} {\bibinfo {title} {Competing electronic orders on kagome lattices at van hove filling},\ }\href {https://doi.org/10.1103/PhysRevB.87.115135} {\bibfield  {journal} {\bibinfo  {journal} {Phys. Rev. B}\ }\textbf {\bibinfo {volume} {87}},\ \bibinfo {pages} {115135} (\bibinfo {year} {2013})}\BibitemShut {NoStop}%
\bibitem [{\citenamefont {Kiesel}\ \emph {et~al.}(2013)\citenamefont {Kiesel}, \citenamefont {Platt},\ and\ \citenamefont {Thomale}}]{kiesel2013unconventional}%
  \BibitemOpen
  \bibfield  {author} {\bibinfo {author} {\bibfnamefont {M.~L.}\ \bibnamefont {Kiesel}}, \bibinfo {author} {\bibfnamefont {C.}~\bibnamefont {Platt}},\ and\ \bibinfo {author} {\bibfnamefont {R.}~\bibnamefont {Thomale}},\ }\bibfield  {title} {\bibinfo {title} {{Unconventional Fermi surface instabilities in the kagome Hubbard model}},\ }\href@noop {} {\bibfield  {journal} {\bibinfo  {journal} {Phys. Rev. Lett.}\ }\textbf {\bibinfo {volume} {110}},\ \bibinfo {pages} {126405} (\bibinfo {year} {2013})}\BibitemShut {NoStop}%
\bibitem [{\citenamefont {Meier}\ \emph {et~al.}(2020)\citenamefont {Meier}, \citenamefont {Du}, \citenamefont {Okamoto}, \citenamefont {Mohanta}, \citenamefont {May}, \citenamefont {McGuire}, \citenamefont {Bridges}, \citenamefont {Samolyuk},\ and\ \citenamefont {Sales}}]{meier2020flat}%
  \BibitemOpen
  \bibfield  {author} {\bibinfo {author} {\bibfnamefont {W.~R.}\ \bibnamefont {Meier}}, \bibinfo {author} {\bibfnamefont {M.-H.}\ \bibnamefont {Du}}, \bibinfo {author} {\bibfnamefont {S.}~\bibnamefont {Okamoto}}, \bibinfo {author} {\bibfnamefont {N.}~\bibnamefont {Mohanta}}, \bibinfo {author} {\bibfnamefont {A.~F.}\ \bibnamefont {May}}, \bibinfo {author} {\bibfnamefont {M.~A.}\ \bibnamefont {McGuire}}, \bibinfo {author} {\bibfnamefont {C.~A.}\ \bibnamefont {Bridges}}, \bibinfo {author} {\bibfnamefont {G.~D.}\ \bibnamefont {Samolyuk}},\ and\ \bibinfo {author} {\bibfnamefont {B.~C.}\ \bibnamefont {Sales}},\ }\bibfield  {title} {\bibinfo {title} {{Flat bands in the CoSn-type compounds}},\ }\href@noop {} {\bibfield  {journal} {\bibinfo  {journal} {Phys. Rev. B}\ }\textbf {\bibinfo {volume} {102}},\ \bibinfo {pages} {075148} (\bibinfo {year} {2020})}\BibitemShut {NoStop}%
\bibitem [{\citenamefont {Wilson}\ and\ \citenamefont {Ortiz}(2024)}]{wilsonReview}%
  \BibitemOpen
  \bibfield  {author} {\bibinfo {author} {\bibfnamefont {S.~D.}\ \bibnamefont {Wilson}}\ and\ \bibinfo {author} {\bibfnamefont {B.~R.}\ \bibnamefont {Ortiz}},\ }\bibfield  {title} {\bibinfo {title} {{AV$_3$Sb$_5$ kagome superconductors}},\ }\href@noop {} {\bibfield  {journal} {\bibinfo  {journal} {Nat. Rev. Mater.}\ ,\ \bibinfo {pages} {9756}} (\bibinfo {year} {2024})}\BibitemShut {NoStop}%
\bibitem [{\citenamefont {Ortiz}\ \emph {et~al.}(2020{\natexlab{a}})\citenamefont {Ortiz}, \citenamefont {Teicher}, \citenamefont {Hu}, \citenamefont {Zuo}, \citenamefont {Sarte}, \citenamefont {Schueller}, \citenamefont {Abeykoon}, \citenamefont {Krogstad}, \citenamefont {Rosenkranz}, \citenamefont {Osborn}, \citenamefont {Seshadri}, \citenamefont {Balents}, \citenamefont {He},\ and\ \citenamefont {Wilson}}]{ortizCsV3Sb5}%
  \BibitemOpen
  \bibfield  {author} {\bibinfo {author} {\bibfnamefont {B.~R.}\ \bibnamefont {Ortiz}}, \bibinfo {author} {\bibfnamefont {S.~M.}\ \bibnamefont {Teicher}}, \bibinfo {author} {\bibfnamefont {Y.}~\bibnamefont {Hu}}, \bibinfo {author} {\bibfnamefont {J.~L.}\ \bibnamefont {Zuo}}, \bibinfo {author} {\bibfnamefont {P.~M.}\ \bibnamefont {Sarte}}, \bibinfo {author} {\bibfnamefont {E.~C.}\ \bibnamefont {Schueller}}, \bibinfo {author} {\bibfnamefont {A.~M.}\ \bibnamefont {Abeykoon}}, \bibinfo {author} {\bibfnamefont {M.~J.}\ \bibnamefont {Krogstad}}, \bibinfo {author} {\bibfnamefont {S.}~\bibnamefont {Rosenkranz}}, \bibinfo {author} {\bibfnamefont {R.}~\bibnamefont {Osborn}}, \bibinfo {author} {\bibfnamefont {R.}~\bibnamefont {Seshadri}}, \bibinfo {author} {\bibfnamefont {L.}~\bibnamefont {Balents}}, \bibinfo {author} {\bibfnamefont {J.}~\bibnamefont {He}},\ and\ \bibinfo {author} {\bibfnamefont {S.~D.}\ \bibnamefont {Wilson}},\ }\bibfield  {title} {\bibinfo {title} {{CsV$_3$Sb$_5$: a $\mathbb{Z}_2$ topological kagome
  metal with a superconducting ground state}},\ }\href@noop {} {\bibfield  {journal} {\bibinfo  {journal} {Phys. Rev. Lett.}\ }\textbf {\bibinfo {volume} {125}},\ \bibinfo {pages} {247002} (\bibinfo {year} {2020}{\natexlab{a}})}\BibitemShut {NoStop}%
\bibitem [{\citenamefont {Ortiz}\ \emph {et~al.}(2020{\natexlab{b}})\citenamefont {Ortiz}, \citenamefont {Kenney}, \citenamefont {Sarte}, \citenamefont {Teicher}, \citenamefont {Seshadri}, \citenamefont {Graf},\ and\ \citenamefont {Wilson}}]{ortiz2020KV3Sb5}%
  \BibitemOpen
  \bibfield  {author} {\bibinfo {author} {\bibfnamefont {B.~R.}\ \bibnamefont {Ortiz}}, \bibinfo {author} {\bibfnamefont {E.}~\bibnamefont {Kenney}}, \bibinfo {author} {\bibfnamefont {P.~M.}\ \bibnamefont {Sarte}}, \bibinfo {author} {\bibfnamefont {S.~M.}\ \bibnamefont {Teicher}}, \bibinfo {author} {\bibfnamefont {R.}~\bibnamefont {Seshadri}}, \bibinfo {author} {\bibfnamefont {M.~J.}\ \bibnamefont {Graf}},\ and\ \bibinfo {author} {\bibfnamefont {S.~D.}\ \bibnamefont {Wilson}},\ }\bibfield  {title} {\bibinfo {title} {{Superconductivity in the $\mathbb{Z}_2$ kagome metal KV$_3$Sb$_5$}},\ }\href@noop {} {\bibfield  {journal} {\bibinfo  {journal} {Phys. Rev. Mater.}\ }\textbf {\bibinfo {volume} {5}},\ \bibinfo {pages} {034801} (\bibinfo {year} {2020}{\natexlab{b}})}\BibitemShut {NoStop}%
\bibitem [{\citenamefont {Yin}\ \emph {et~al.}(2021)\citenamefont {Yin}, \citenamefont {Tu}, \citenamefont {Gong}, \citenamefont {Fu}, \citenamefont {Yan},\ and\ \citenamefont {Lei}}]{RbV3Sb5SC}%
  \BibitemOpen
  \bibfield  {author} {\bibinfo {author} {\bibfnamefont {Q.}~\bibnamefont {Yin}}, \bibinfo {author} {\bibfnamefont {Z.}~\bibnamefont {Tu}}, \bibinfo {author} {\bibfnamefont {C.}~\bibnamefont {Gong}}, \bibinfo {author} {\bibfnamefont {Y.}~\bibnamefont {Fu}}, \bibinfo {author} {\bibfnamefont {S.}~\bibnamefont {Yan}},\ and\ \bibinfo {author} {\bibfnamefont {H.}~\bibnamefont {Lei}},\ }\bibfield  {title} {\bibinfo {title} {{Superconductivity and normal-state properties of kagome metal RbV$_3$Sb$_5$ single crystals}},\ }\href@noop {} {\bibfield  {journal} {\bibinfo  {journal} {Chin. Phys. Lett.}\ }\textbf {\bibinfo {volume} {38}},\ \bibinfo {pages} {037403} (\bibinfo {year} {2021})}\BibitemShut {NoStop}%
\bibitem [{\citenamefont {Ortiz}\ \emph {et~al.}(2021)\citenamefont {Ortiz}, \citenamefont {Teicher}, \citenamefont {Kautzsch}, \citenamefont {Sarte}, \citenamefont {Ratcliff}, \citenamefont {Harter}, \citenamefont {Ruff}, \citenamefont {Seshadri},\ and\ \citenamefont {Wilson}}]{ortiz2021fermi}%
  \BibitemOpen
  \bibfield  {author} {\bibinfo {author} {\bibfnamefont {B.~R.}\ \bibnamefont {Ortiz}}, \bibinfo {author} {\bibfnamefont {S.~M.}\ \bibnamefont {Teicher}}, \bibinfo {author} {\bibfnamefont {L.}~\bibnamefont {Kautzsch}}, \bibinfo {author} {\bibfnamefont {P.~M.}\ \bibnamefont {Sarte}}, \bibinfo {author} {\bibfnamefont {N.}~\bibnamefont {Ratcliff}}, \bibinfo {author} {\bibfnamefont {J.}~\bibnamefont {Harter}}, \bibinfo {author} {\bibfnamefont {J.~P.}\ \bibnamefont {Ruff}}, \bibinfo {author} {\bibfnamefont {R.}~\bibnamefont {Seshadri}},\ and\ \bibinfo {author} {\bibfnamefont {S.~D.}\ \bibnamefont {Wilson}},\ }\bibfield  {title} {\bibinfo {title} {{Fermi surface mapping and the nature of charge-density-wave order in the kagome superconductor CsV$_3$Sb$_5$}},\ }\href@noop {} {\bibfield  {journal} {\bibinfo  {journal} {Phys. Rev. X}\ }\textbf {\bibinfo {volume} {11}},\ \bibinfo {pages} {041030} (\bibinfo {year} {2021})}\BibitemShut {NoStop}%
\bibitem [{\citenamefont {Zhao}\ \emph {et~al.}(2021)\citenamefont {Zhao}, \citenamefont {Li}, \citenamefont {Ortiz}, \citenamefont {Teicher}, \citenamefont {Park}, \citenamefont {Ye}, \citenamefont {Wang}, \citenamefont {Balents}, \citenamefont {Wilson},\ and\ \citenamefont {Zeljkovic}}]{zhao2021cascade}%
  \BibitemOpen
  \bibfield  {author} {\bibinfo {author} {\bibfnamefont {H.}~\bibnamefont {Zhao}}, \bibinfo {author} {\bibfnamefont {H.}~\bibnamefont {Li}}, \bibinfo {author} {\bibfnamefont {B.~R.}\ \bibnamefont {Ortiz}}, \bibinfo {author} {\bibfnamefont {S.~M.}\ \bibnamefont {Teicher}}, \bibinfo {author} {\bibfnamefont {T.}~\bibnamefont {Park}}, \bibinfo {author} {\bibfnamefont {M.}~\bibnamefont {Ye}}, \bibinfo {author} {\bibfnamefont {Z.}~\bibnamefont {Wang}}, \bibinfo {author} {\bibfnamefont {L.}~\bibnamefont {Balents}}, \bibinfo {author} {\bibfnamefont {S.~D.}\ \bibnamefont {Wilson}},\ and\ \bibinfo {author} {\bibfnamefont {I.}~\bibnamefont {Zeljkovic}},\ }\bibfield  {title} {\bibinfo {title} {{Cascade of correlated electron states in the kagome superconductor CsV$_3$Sb$_5$}},\ }\href@noop {} {\bibfield  {journal} {\bibinfo  {journal} {Nature}\ }\textbf {\bibinfo {volume} {599}},\ \bibinfo {pages} {216} (\bibinfo {year} {2021})}\BibitemShut {NoStop}%
\bibitem [{\citenamefont {Hu}\ \emph {et~al.}(2022)\citenamefont {Hu}, \citenamefont {Wu}, \citenamefont {Ortiz}, \citenamefont {Han}, \citenamefont {Plumb}, \citenamefont {Wilson}, \citenamefont {Schnyder}, \citenamefont {Shi} \emph {et~al.}}]{hu2022coexistence}%
  \BibitemOpen
  \bibfield  {author} {\bibinfo {author} {\bibfnamefont {Y.}~\bibnamefont {Hu}}, \bibinfo {author} {\bibfnamefont {X.}~\bibnamefont {Wu}}, \bibinfo {author} {\bibfnamefont {B.~R.}\ \bibnamefont {Ortiz}}, \bibinfo {author} {\bibfnamefont {X.}~\bibnamefont {Han}}, \bibinfo {author} {\bibfnamefont {N.~C.}\ \bibnamefont {Plumb}}, \bibinfo {author} {\bibfnamefont {S.~D.}\ \bibnamefont {Wilson}}, \bibinfo {author} {\bibfnamefont {A.~P.}\ \bibnamefont {Schnyder}}, \bibinfo {author} {\bibfnamefont {M.}~\bibnamefont {Shi}}, \emph {et~al.},\ }\bibfield  {title} {\bibinfo {title} {{Coexistence of trihexagonal and star-of-David pattern in the charge density wave of the kagome superconductor \textit{A}V$_3$Sb$_5$}},\ }\href@noop {} {\bibfield  {journal} {\bibinfo  {journal} {Phys. Rev. B}\ }\textbf {\bibinfo {volume} {106}},\ \bibinfo {pages} {L241106} (\bibinfo {year} {2022})}\BibitemShut {NoStop}%
\bibitem [{\citenamefont {Kang}\ \emph {et~al.}(2023)\citenamefont {Kang}, \citenamefont {Fang}, \citenamefont {Yoo}, \citenamefont {Ortiz}, \citenamefont {Oey}, \citenamefont {Choi}, \citenamefont {Ryu}, \citenamefont {Kim}, \citenamefont {Jozwiak}, \citenamefont {Bostwick} \emph {et~al.}}]{kang2022microscopic}%
  \BibitemOpen
  \bibfield  {author} {\bibinfo {author} {\bibfnamefont {M.}~\bibnamefont {Kang}}, \bibinfo {author} {\bibfnamefont {S.}~\bibnamefont {Fang}}, \bibinfo {author} {\bibfnamefont {J.}~\bibnamefont {Yoo}}, \bibinfo {author} {\bibfnamefont {B.~R.}\ \bibnamefont {Ortiz}}, \bibinfo {author} {\bibfnamefont {Y.~M.}\ \bibnamefont {Oey}}, \bibinfo {author} {\bibfnamefont {J.}~\bibnamefont {Choi}}, \bibinfo {author} {\bibfnamefont {S.~H.}\ \bibnamefont {Ryu}}, \bibinfo {author} {\bibfnamefont {J.}~\bibnamefont {Kim}}, \bibinfo {author} {\bibfnamefont {C.}~\bibnamefont {Jozwiak}}, \bibinfo {author} {\bibfnamefont {A.}~\bibnamefont {Bostwick}}, \emph {et~al.},\ }\bibfield  {title} {\bibinfo {title} {Charge order landscape and competition with superconductivity in kagome metals},\ }\href@noop {} {\bibfield  {journal} {\bibinfo  {journal} {Nat. Mater.}\ }\textbf {\bibinfo {volume} {22}},\ \bibinfo {pages} {186} (\bibinfo {year} {2023})}\BibitemShut {NoStop}%
\bibitem [{\citenamefont {Jiang}\ \emph {et~al.}(2021)\citenamefont {Jiang}, \citenamefont {Yin}, \citenamefont {Denner}, \citenamefont {Shumiya}, \citenamefont {Ortiz}, \citenamefont {Xu}, \citenamefont {Guguchia}, \citenamefont {He}, \citenamefont {Hossain}, \citenamefont {Liu} \emph {et~al.}}]{jiang2021unconventional}%
  \BibitemOpen
  \bibfield  {author} {\bibinfo {author} {\bibfnamefont {Y.-X.}\ \bibnamefont {Jiang}}, \bibinfo {author} {\bibfnamefont {J.-X.}\ \bibnamefont {Yin}}, \bibinfo {author} {\bibfnamefont {M.~M.}\ \bibnamefont {Denner}}, \bibinfo {author} {\bibfnamefont {N.}~\bibnamefont {Shumiya}}, \bibinfo {author} {\bibfnamefont {B.~R.}\ \bibnamefont {Ortiz}}, \bibinfo {author} {\bibfnamefont {G.}~\bibnamefont {Xu}}, \bibinfo {author} {\bibfnamefont {Z.}~\bibnamefont {Guguchia}}, \bibinfo {author} {\bibfnamefont {J.}~\bibnamefont {He}}, \bibinfo {author} {\bibfnamefont {M.~S.}\ \bibnamefont {Hossain}}, \bibinfo {author} {\bibfnamefont {X.}~\bibnamefont {Liu}}, \emph {et~al.},\ }\bibfield  {title} {\bibinfo {title} {{Unconventional chiral charge order in kagome superconductor KV$_3$Sb$_5$}},\ }\href@noop {} {\bibfield  {journal} {\bibinfo  {journal} {Nat. Mater.}\ }\textbf {\bibinfo {volume} {20}},\ \bibinfo {pages} {1353} (\bibinfo {year} {2021})}\BibitemShut {NoStop}%
\bibitem [{\citenamefont {Liu}\ \emph {et~al.}()\citenamefont {Liu}, \citenamefont {Liu}, \citenamefont {Bao}, \citenamefont {Yang}, \citenamefont {Ji}, \citenamefont {Liu}, \citenamefont {Xu}, \citenamefont {Yang}, \citenamefont {Chai}, \citenamefont {Lu} \emph {et~al.}}]{liu2023superconductivity}%
  \BibitemOpen
  \bibfield  {author} {\bibinfo {author} {\bibfnamefont {Y.}~\bibnamefont {Liu}}, \bibinfo {author} {\bibfnamefont {Z.-Y.}\ \bibnamefont {Liu}}, \bibinfo {author} {\bibfnamefont {J.-K.}\ \bibnamefont {Bao}}, \bibinfo {author} {\bibfnamefont {P.-T.}\ \bibnamefont {Yang}}, \bibinfo {author} {\bibfnamefont {L.-W.}\ \bibnamefont {Ji}}, \bibinfo {author} {\bibfnamefont {J.-Y.}\ \bibnamefont {Liu}}, \bibinfo {author} {\bibfnamefont {C.-C.}\ \bibnamefont {Xu}}, \bibinfo {author} {\bibfnamefont {W.-Z.}\ \bibnamefont {Yang}}, \bibinfo {author} {\bibfnamefont {W.-L.}\ \bibnamefont {Chai}}, \bibinfo {author} {\bibfnamefont {J.-Y.}\ \bibnamefont {Lu}}, \emph {et~al.},\ }\bibfield  {title} {\bibinfo {title} {{Superconductivity emerged from density-wave order in a kagome bad metal}},\ }\href@noop {} {\bibinfo  {journal} {\textbf{2023}, arXiv:2309.13514 [cond-mat.supr-con]. arXiv.org e-Print archive. //https://arxiv.org/pdf/2309.13514 (Accessed 6-25-2024)}\ }\BibitemShut {NoStop}%
\bibitem [{\citenamefont {Wang}\ \emph {et~al.}(2022)\citenamefont {Wang}, \citenamefont {McCandless}, \citenamefont {Wang}, \citenamefont {Thanabalasingam}, \citenamefont {Wu}, \citenamefont {Bouwmeester}, \citenamefont {Van Der~Zant}, \citenamefont {Ali},\ and\ \citenamefont {Chan}}]{wang2022electronic}%
  \BibitemOpen
\bibfield  {journal} {  }\bibfield  {author} {\bibinfo {author} {\bibfnamefont {Y.}~\bibnamefont {Wang}}, \bibinfo {author} {\bibfnamefont {G.~T.}\ \bibnamefont {McCandless}}, \bibinfo {author} {\bibfnamefont {X.}~\bibnamefont {Wang}}, \bibinfo {author} {\bibfnamefont {K.}~\bibnamefont {Thanabalasingam}}, \bibinfo {author} {\bibfnamefont {H.}~\bibnamefont {Wu}}, \bibinfo {author} {\bibfnamefont {D.}~\bibnamefont {Bouwmeester}}, \bibinfo {author} {\bibfnamefont {H.~S.}\ \bibnamefont {Van Der~Zant}}, \bibinfo {author} {\bibfnamefont {M.~N.}\ \bibnamefont {Ali}},\ and\ \bibinfo {author} {\bibfnamefont {J.~Y.}\ \bibnamefont {Chan}},\ }\bibfield  {title} {\bibinfo {title} {{Electronic properties and phase transition in the kagome metal Yb$_{0.5}$Co$_3$Ge$_3$}},\ }\href@noop {} {\bibfield  {journal} {\bibinfo  {journal} {Chemistry of Materials}\ }\textbf {\bibinfo {volume} {34}},\ \bibinfo {pages} {7337} (\bibinfo {year} {2022})}\BibitemShut {NoStop}%
\bibitem [{\citenamefont {Peng}\ \emph {et~al.}(2021)\citenamefont {Peng}, \citenamefont {Han}, \citenamefont {Pokharel}, \citenamefont {Shen}, \citenamefont {Li}, \citenamefont {Hashimoto}, \citenamefont {Lu}, \citenamefont {Ortiz}, \citenamefont {Luo}, \citenamefont {Li}, \citenamefont {Guo}, \citenamefont {Wang}, \citenamefont {Cui}, \citenamefont {Sun}, \citenamefont {Qiao}, \citenamefont {Wilson},\ and\ \citenamefont {He}}]{PhysRevLett.127.266401}%
  \BibitemOpen
  \bibfield  {author} {\bibinfo {author} {\bibfnamefont {S.}~\bibnamefont {Peng}}, \bibinfo {author} {\bibfnamefont {Y.}~\bibnamefont {Han}}, \bibinfo {author} {\bibfnamefont {G.}~\bibnamefont {Pokharel}}, \bibinfo {author} {\bibfnamefont {J.}~\bibnamefont {Shen}}, \bibinfo {author} {\bibfnamefont {Z.}~\bibnamefont {Li}}, \bibinfo {author} {\bibfnamefont {M.}~\bibnamefont {Hashimoto}}, \bibinfo {author} {\bibfnamefont {D.}~\bibnamefont {Lu}}, \bibinfo {author} {\bibfnamefont {B.~R.}\ \bibnamefont {Ortiz}}, \bibinfo {author} {\bibfnamefont {Y.}~\bibnamefont {Luo}}, \bibinfo {author} {\bibfnamefont {H.}~\bibnamefont {Li}}, \bibinfo {author} {\bibfnamefont {M.}~\bibnamefont {Guo}}, \bibinfo {author} {\bibfnamefont {B.}~\bibnamefont {Wang}}, \bibinfo {author} {\bibfnamefont {S.}~\bibnamefont {Cui}}, \bibinfo {author} {\bibfnamefont {Z.}~\bibnamefont {Sun}}, \bibinfo {author} {\bibfnamefont {Z.}~\bibnamefont {Qiao}}, \bibinfo {author} {\bibfnamefont {S.~D.}\ \bibnamefont {Wilson}},\ and\ \bibinfo {author}
  {\bibfnamefont {J.}~\bibnamefont {He}},\ }\bibfield  {title} {\bibinfo {title} {Realizing kagome band structure in two-dimensional kagome surface states of \text{$R{\mathrm{V}}_{6}{\mathrm{Sn}}_{6}$ ($R=\mathrm{Gd}$, Ho)}},\ }\href {https://doi.org/10.1103/PhysRevLett.127.266401} {\bibfield  {journal} {\bibinfo  {journal} {Phys. Rev. Lett.}\ }\textbf {\bibinfo {volume} {127}},\ \bibinfo {pages} {266401} (\bibinfo {year} {2021})}\BibitemShut {NoStop}%
\bibitem [{\citenamefont {Wang}\ \emph {et~al.}(2021)\citenamefont {Wang}, \citenamefont {Neubauer}, \citenamefont {Duan}, \citenamefont {Yin}, \citenamefont {Fujitsu}, \citenamefont {Hosono}, \citenamefont {Ye}, \citenamefont {Zhang}, \citenamefont {Chi}, \citenamefont {Krycka}, \citenamefont {Lei},\ and\ \citenamefont {Dai}}]{PhysRevB.103.014416}%
  \BibitemOpen
  \bibfield  {author} {\bibinfo {author} {\bibfnamefont {Q.}~\bibnamefont {Wang}}, \bibinfo {author} {\bibfnamefont {K.~J.}\ \bibnamefont {Neubauer}}, \bibinfo {author} {\bibfnamefont {C.}~\bibnamefont {Duan}}, \bibinfo {author} {\bibfnamefont {Q.}~\bibnamefont {Yin}}, \bibinfo {author} {\bibfnamefont {S.}~\bibnamefont {Fujitsu}}, \bibinfo {author} {\bibfnamefont {H.}~\bibnamefont {Hosono}}, \bibinfo {author} {\bibfnamefont {F.}~\bibnamefont {Ye}}, \bibinfo {author} {\bibfnamefont {R.}~\bibnamefont {Zhang}}, \bibinfo {author} {\bibfnamefont {S.}~\bibnamefont {Chi}}, \bibinfo {author} {\bibfnamefont {K.}~\bibnamefont {Krycka}}, \bibinfo {author} {\bibfnamefont {H.}~\bibnamefont {Lei}},\ and\ \bibinfo {author} {\bibfnamefont {P.}~\bibnamefont {Dai}},\ }\bibfield  {title} {\bibinfo {title} {{Field-induced topological Hall effect and double-fan spin structure with a $c$-axis component in the metallic kagome antiferromagnetic compound YMn$_6$Sn$_6$}},\ }\href {https://doi.org/10.1103/PhysRevB.103.014416}
  {\bibfield  {journal} {\bibinfo  {journal} {Phys. Rev. B}\ }\textbf {\bibinfo {volume} {103}},\ \bibinfo {pages} {014416} (\bibinfo {year} {2021})}\BibitemShut {NoStop}%
\bibitem [{\citenamefont {Pokharel}\ \emph {et~al.}(2021)\citenamefont {Pokharel}, \citenamefont {Teicher}, \citenamefont {Ortiz}, \citenamefont {Sarte}, \citenamefont {Wu}, \citenamefont {Peng}, \citenamefont {He}, \citenamefont {Seshadri},\ and\ \citenamefont {Wilson}}]{PhysRevB.104.235139}%
  \BibitemOpen
  \bibfield  {author} {\bibinfo {author} {\bibfnamefont {G.}~\bibnamefont {Pokharel}}, \bibinfo {author} {\bibfnamefont {S.~M.~L.}\ \bibnamefont {Teicher}}, \bibinfo {author} {\bibfnamefont {B.~R.}\ \bibnamefont {Ortiz}}, \bibinfo {author} {\bibfnamefont {P.~M.}\ \bibnamefont {Sarte}}, \bibinfo {author} {\bibfnamefont {G.}~\bibnamefont {Wu}}, \bibinfo {author} {\bibfnamefont {S.}~\bibnamefont {Peng}}, \bibinfo {author} {\bibfnamefont {J.}~\bibnamefont {He}}, \bibinfo {author} {\bibfnamefont {R.}~\bibnamefont {Seshadri}},\ and\ \bibinfo {author} {\bibfnamefont {S.~D.}\ \bibnamefont {Wilson}},\ }\bibfield  {title} {\bibinfo {title} {{Electronic properties of the topological kagome metals YV$_6$Sn$_6$ and GdV$_6$Sn$_6$}},\ }\href {https://doi.org/10.1103/PhysRevB.104.235139} {\bibfield  {journal} {\bibinfo  {journal} {Phys. Rev. B}\ }\textbf {\bibinfo {volume} {104}},\ \bibinfo {pages} {235139} (\bibinfo {year} {2021})}\BibitemShut {NoStop}%
\bibitem [{\citenamefont {Pokharel}\ \emph {et~al.}(2022)\citenamefont {Pokharel}, \citenamefont {Ortiz}, \citenamefont {Chamorro}, \citenamefont {Sarte}, \citenamefont {Kautzsch}, \citenamefont {Wu}, \citenamefont {Ruff},\ and\ \citenamefont {Wilson}}]{PhysRevMaterials104202}%
  \BibitemOpen
  \bibfield  {author} {\bibinfo {author} {\bibfnamefont {G.}~\bibnamefont {Pokharel}}, \bibinfo {author} {\bibfnamefont {B.}~\bibnamefont {Ortiz}}, \bibinfo {author} {\bibfnamefont {J.}~\bibnamefont {Chamorro}}, \bibinfo {author} {\bibfnamefont {P.}~\bibnamefont {Sarte}}, \bibinfo {author} {\bibfnamefont {L.}~\bibnamefont {Kautzsch}}, \bibinfo {author} {\bibfnamefont {G.}~\bibnamefont {Wu}}, \bibinfo {author} {\bibfnamefont {J.}~\bibnamefont {Ruff}},\ and\ \bibinfo {author} {\bibfnamefont {S.~D.}\ \bibnamefont {Wilson}},\ }\bibfield  {title} {\bibinfo {title} {{Highly anisotropic magnetism in the vanadium-based kagome metal TbV$_6$Sn$_6$}},\ }\href {https://doi.org/10.1103/PhysRevMaterials.6.104202} {\bibfield  {journal} {\bibinfo  {journal} {Phys. Rev. Mater.}\ }\textbf {\bibinfo {volume} {6}},\ \bibinfo {pages} {104202} (\bibinfo {year} {2022})}\BibitemShut {NoStop}%
\bibitem [{\citenamefont {Rosenberg}\ \emph {et~al.}(2022)\citenamefont {Rosenberg}, \citenamefont {DeStefano}, \citenamefont {Guo}, \citenamefont {Oh}, \citenamefont {Hashimoto}, \citenamefont {Lu}, \citenamefont {Birgeneau}, \citenamefont {Lee}, \citenamefont {Ke}, \citenamefont {Yi},\ and\ \citenamefont {Chu}}]{PhysRevB.106.115139}%
  \BibitemOpen
  \bibfield  {author} {\bibinfo {author} {\bibfnamefont {E.}~\bibnamefont {Rosenberg}}, \bibinfo {author} {\bibfnamefont {J.~M.}\ \bibnamefont {DeStefano}}, \bibinfo {author} {\bibfnamefont {Y.}~\bibnamefont {Guo}}, \bibinfo {author} {\bibfnamefont {J.~S.}\ \bibnamefont {Oh}}, \bibinfo {author} {\bibfnamefont {M.}~\bibnamefont {Hashimoto}}, \bibinfo {author} {\bibfnamefont {D.}~\bibnamefont {Lu}}, \bibinfo {author} {\bibfnamefont {R.~J.}\ \bibnamefont {Birgeneau}}, \bibinfo {author} {\bibfnamefont {Y.}~\bibnamefont {Lee}}, \bibinfo {author} {\bibfnamefont {L.}~\bibnamefont {Ke}}, \bibinfo {author} {\bibfnamefont {M.}~\bibnamefont {Yi}},\ and\ \bibinfo {author} {\bibfnamefont {J.-H.}\ \bibnamefont {Chu}},\ }\bibfield  {title} {\bibinfo {title} {{Uniaxial ferromagnetism in the kagome metal TbV$_6$Sn$_6$}},\ }\href {https://doi.org/10.1103/PhysRevB.106.115139} {\bibfield  {journal} {\bibinfo  {journal} {Phys. Rev. B}\ }\textbf {\bibinfo {volume} {106}},\ \bibinfo {pages} {115139} (\bibinfo {year}
  {2022})}\BibitemShut {NoStop}%
\bibitem [{\citenamefont {Ghimire}\ \emph {et~al.}(2020)\citenamefont {Ghimire}, \citenamefont {Dally}, \citenamefont {Poudel}, \citenamefont {Jones}, \citenamefont {Michel}, \citenamefont {Magar}, \citenamefont {Bleuel}, \citenamefont {McGuire}, \citenamefont {Jiang}, \citenamefont {Mitchell}, \citenamefont {Lynn},\ and\ \citenamefont {Mazin}}]{sciadv_abe2680}%
  \BibitemOpen
  \bibfield  {author} {\bibinfo {author} {\bibfnamefont {N.~J.}\ \bibnamefont {Ghimire}}, \bibinfo {author} {\bibfnamefont {R.~L.}\ \bibnamefont {Dally}}, \bibinfo {author} {\bibfnamefont {L.}~\bibnamefont {Poudel}}, \bibinfo {author} {\bibfnamefont {D.~C.}\ \bibnamefont {Jones}}, \bibinfo {author} {\bibfnamefont {D.}~\bibnamefont {Michel}}, \bibinfo {author} {\bibfnamefont {N.~T.}\ \bibnamefont {Magar}}, \bibinfo {author} {\bibfnamefont {M.}~\bibnamefont {Bleuel}}, \bibinfo {author} {\bibfnamefont {M.~A.}\ \bibnamefont {McGuire}}, \bibinfo {author} {\bibfnamefont {J.~S.}\ \bibnamefont {Jiang}}, \bibinfo {author} {\bibfnamefont {J.~F.}\ \bibnamefont {Mitchell}}, \bibinfo {author} {\bibfnamefont {J.~W.}\ \bibnamefont {Lynn}},\ and\ \bibinfo {author} {\bibfnamefont {I.~I.}\ \bibnamefont {Mazin}},\ }\bibfield  {title} {\bibinfo {title} {Competing magnetic phases and fluctuation-driven scalar spin chirality in the kagome metal \text{YMn$_6$Sn$_6$}},\ }\href {https://doi.org/10.1126/sciadv.abe2680} {\bibfield
  {journal} {\bibinfo  {journal} {Science Advances}\ }\textbf {\bibinfo {volume} {6}},\ \bibinfo {pages} {eabe2680} (\bibinfo {year} {2020})}\BibitemShut {NoStop}%
\bibitem [{\citenamefont {Arachchige}\ \emph {et~al.}(2022)\citenamefont {Arachchige}, \citenamefont {Meier}, \citenamefont {Marshall}, \citenamefont {Matsuoka}, \citenamefont {Xue}, \citenamefont {McGuire}, \citenamefont {Hermann}, \citenamefont {Cao},\ and\ \citenamefont {Mandrus}}]{PhysRevLett.129.216402}%
  \BibitemOpen
  \bibfield  {author} {\bibinfo {author} {\bibfnamefont {H.~W.~S.}\ \bibnamefont {Arachchige}}, \bibinfo {author} {\bibfnamefont {W.~R.}\ \bibnamefont {Meier}}, \bibinfo {author} {\bibfnamefont {M.}~\bibnamefont {Marshall}}, \bibinfo {author} {\bibfnamefont {T.}~\bibnamefont {Matsuoka}}, \bibinfo {author} {\bibfnamefont {R.}~\bibnamefont {Xue}}, \bibinfo {author} {\bibfnamefont {M.~A.}\ \bibnamefont {McGuire}}, \bibinfo {author} {\bibfnamefont {R.~P.}\ \bibnamefont {Hermann}}, \bibinfo {author} {\bibfnamefont {H.}~\bibnamefont {Cao}},\ and\ \bibinfo {author} {\bibfnamefont {D.}~\bibnamefont {Mandrus}},\ }\bibfield  {title} {\bibinfo {title} {Charge density wave in kagome lattice intermetallic \text{${\mathrm{ScV}}_{6}{\mathrm{Sn}}_{6}$}},\ }\href {https://doi.org/10.1103/PhysRevLett.129.216402} {\bibfield  {journal} {\bibinfo  {journal} {Phys. Rev. Lett.}\ }\textbf {\bibinfo {volume} {129}},\ \bibinfo {pages} {216402} (\bibinfo {year} {2022})}\BibitemShut {NoStop}%
\bibitem [{\citenamefont {Yin}\ \emph {et~al.}(2020)\citenamefont {Yin}, \citenamefont {Ma}, \citenamefont {Cochran}, \citenamefont {Xu}, \citenamefont {Zhang}, \citenamefont {Tien}, \citenamefont {Shumiya}, \citenamefont {Cheng}, \citenamefont {Jiang}, \citenamefont {Lian}, \citenamefont {Song}, \citenamefont {Chang}, \citenamefont {Belopolski}, \citenamefont {Multer}, \citenamefont {Litskevich}, \citenamefont {Cheng}, \citenamefont {Yang}, \citenamefont {Swidler}, \citenamefont {Zhou}, \citenamefont {Lin}, \citenamefont {Neupert}, \citenamefont {Wang}, \citenamefont {Yao}, \citenamefont {Chang}, \citenamefont {Jia},\ and\ \citenamefont {Zahid~Hasan}}]{Yin_2020}%
  \BibitemOpen
  \bibfield  {author} {\bibinfo {author} {\bibfnamefont {J.-X.}\ \bibnamefont {Yin}}, \bibinfo {author} {\bibfnamefont {W.}~\bibnamefont {Ma}}, \bibinfo {author} {\bibfnamefont {T.~A.}\ \bibnamefont {Cochran}}, \bibinfo {author} {\bibfnamefont {X.}~\bibnamefont {Xu}}, \bibinfo {author} {\bibfnamefont {S.~S.}\ \bibnamefont {Zhang}}, \bibinfo {author} {\bibfnamefont {H.-J.}\ \bibnamefont {Tien}}, \bibinfo {author} {\bibfnamefont {N.}~\bibnamefont {Shumiya}}, \bibinfo {author} {\bibfnamefont {G.}~\bibnamefont {Cheng}}, \bibinfo {author} {\bibfnamefont {K.}~\bibnamefont {Jiang}}, \bibinfo {author} {\bibfnamefont {B.}~\bibnamefont {Lian}}, \bibinfo {author} {\bibfnamefont {Z.}~\bibnamefont {Song}}, \bibinfo {author} {\bibfnamefont {G.}~\bibnamefont {Chang}}, \bibinfo {author} {\bibfnamefont {I.}~\bibnamefont {Belopolski}}, \bibinfo {author} {\bibfnamefont {D.}~\bibnamefont {Multer}}, \bibinfo {author} {\bibfnamefont {M.}~\bibnamefont {Litskevich}}, \bibinfo {author} {\bibfnamefont {Z.-J.}\ \bibnamefont {Cheng}},
  \bibinfo {author} {\bibfnamefont {X.~P.}\ \bibnamefont {Yang}}, \bibinfo {author} {\bibfnamefont {B.}~\bibnamefont {Swidler}}, \bibinfo {author} {\bibfnamefont {H.}~\bibnamefont {Zhou}}, \bibinfo {author} {\bibfnamefont {H.}~\bibnamefont {Lin}}, \bibinfo {author} {\bibfnamefont {T.}~\bibnamefont {Neupert}}, \bibinfo {author} {\bibfnamefont {Z.}~\bibnamefont {Wang}}, \bibinfo {author} {\bibfnamefont {N.}~\bibnamefont {Yao}}, \bibinfo {author} {\bibfnamefont {T.-R.}\ \bibnamefont {Chang}}, \bibinfo {author} {\bibfnamefont {S.}~\bibnamefont {Jia}},\ and\ \bibinfo {author} {\bibfnamefont {M.}~\bibnamefont {Zahid~Hasan}},\ }\bibfield  {title} {\bibinfo {title} {Quantum-limit chern topological magnetism in \text{TbMn$_6$Sn$_6$}},\ }\href {https://doi.org/10.1038/s41586-020-2482-7} {\bibfield  {journal} {\bibinfo  {journal} {Nature}\ }\textbf {\bibinfo {volume} {583}},\ \bibinfo {pages} {533} (\bibinfo {year} {2020})}\BibitemShut {NoStop}%
\bibitem [{\citenamefont {Zhang}\ \emph {et~al.}(2022)\citenamefont {Zhang}, \citenamefont {Liu}, \citenamefont {Cui}, \citenamefont {Guo}, \citenamefont {Wang}, \citenamefont {Shi}, \citenamefont {Zhang}, \citenamefont {Wang}, \citenamefont {Dong}, \citenamefont {Sun}, \citenamefont {Dun},\ and\ \citenamefont {Cheng}}]{PhysRevMaterials.6.105001}%
  \BibitemOpen
  \bibfield  {author} {\bibinfo {author} {\bibfnamefont {X.}~\bibnamefont {Zhang}}, \bibinfo {author} {\bibfnamefont {Z.}~\bibnamefont {Liu}}, \bibinfo {author} {\bibfnamefont {Q.}~\bibnamefont {Cui}}, \bibinfo {author} {\bibfnamefont {Q.}~\bibnamefont {Guo}}, \bibinfo {author} {\bibfnamefont {N.}~\bibnamefont {Wang}}, \bibinfo {author} {\bibfnamefont {L.}~\bibnamefont {Shi}}, \bibinfo {author} {\bibfnamefont {H.}~\bibnamefont {Zhang}}, \bibinfo {author} {\bibfnamefont {W.}~\bibnamefont {Wang}}, \bibinfo {author} {\bibfnamefont {X.}~\bibnamefont {Dong}}, \bibinfo {author} {\bibfnamefont {J.}~\bibnamefont {Sun}}, \bibinfo {author} {\bibfnamefont {Z.}~\bibnamefont {Dun}},\ and\ \bibinfo {author} {\bibfnamefont {J.}~\bibnamefont {Cheng}},\ }\bibfield  {title} {\bibinfo {title} {{Electronic and magnetic properties of intermetallic kagome magnets \textit{R}V$_6$Sn$_6$ (\textit{R}: Tb--Tm)}},\ }\href {https://doi.org/10.1103/PhysRevMaterials.6.105001} {\bibfield  {journal} {\bibinfo  {journal} {Phys. Rev. Mater.}\
  }\textbf {\bibinfo {volume} {6}},\ \bibinfo {pages} {105001} (\bibinfo {year} {2022})}\BibitemShut {NoStop}%
\bibitem [{\citenamefont {Lee}\ and\ \citenamefont {Mun}(2022)}]{PhysRevMaterials.6.083401}%
  \BibitemOpen
  \bibfield  {author} {\bibinfo {author} {\bibfnamefont {J.}~\bibnamefont {Lee}}\ and\ \bibinfo {author} {\bibfnamefont {E.}~\bibnamefont {Mun}},\ }\bibfield  {title} {\bibinfo {title} {Anisotropic magnetic property of single crystals \text{$R{\mathrm{V}}_{6}{\mathrm{Sn}}_{6}$ $(R=\mathrm{Y}, \mathrm{Gd}\text{\ensuremath{-}}\mathrm{Tm}, \mathrm{Lu})$}},\ }\href {https://doi.org/10.1103/PhysRevMaterials.6.083401} {\bibfield  {journal} {\bibinfo  {journal} {Phys. Rev. Mater.}\ }\textbf {\bibinfo {volume} {6}},\ \bibinfo {pages} {083401} (\bibinfo {year} {2022})}\BibitemShut {NoStop}%
\bibitem [{\citenamefont {Shao-ying}\ \emph {et~al.}(2001)\citenamefont {Shao-ying}, \citenamefont {Peng}, \citenamefont {Run-wei}, \citenamefont {Sun Ji-rong}, \citenamefont {Hong-wei},\ and\ \citenamefont {Bao-gen}}]{ZhangShao-ying_2001}%
  \BibitemOpen
  \bibfield  {author} {\bibinfo {author} {\bibfnamefont {Z.}~\bibnamefont {Shao-ying}}, \bibinfo {author} {\bibfnamefont {Z.}~\bibnamefont {Peng}}, \bibinfo {author} {\bibfnamefont {L.}~\bibnamefont {Run-wei}}, \bibinfo {author} {\bibfnamefont {C.~Z.-h.}\ \bibnamefont {Sun Ji-rong}}, \bibinfo {author} {\bibfnamefont {Z.}~\bibnamefont {Hong-wei}},\ and\ \bibinfo {author} {\bibfnamefont {S.}~\bibnamefont {Bao-gen}},\ }\bibfield  {title} {\bibinfo {title} {Structure, magnetic properties and giant magnetoresistance of \text{YMn$_6$Sn$_{6-x}$Ga$_x$ (x = 0-0.6)} compounds},\ }\href {https://doi.org/10.1088/1009-1963/10/4/318} {\bibfield  {journal} {\bibinfo  {journal} {Chin. Phys.}\ }\textbf {\bibinfo {volume} {10}},\ \bibinfo {pages} {345} (\bibinfo {year} {2001})}\BibitemShut {NoStop}%
\bibitem [{\citenamefont {Ma}\ \emph {et~al.}(2021)\citenamefont {Ma}, \citenamefont {Xu}, \citenamefont {Yin}, \citenamefont {Yang}, \citenamefont {Zhou}, \citenamefont {Cheng}, \citenamefont {Huang}, \citenamefont {Qu}, \citenamefont {Wang}, \citenamefont {Hasan},\ and\ \citenamefont {Jia}}]{PhysRevLett.126.246602}%
  \BibitemOpen
  \bibfield  {author} {\bibinfo {author} {\bibfnamefont {W.}~\bibnamefont {Ma}}, \bibinfo {author} {\bibfnamefont {X.}~\bibnamefont {Xu}}, \bibinfo {author} {\bibfnamefont {J.-X.}\ \bibnamefont {Yin}}, \bibinfo {author} {\bibfnamefont {H.}~\bibnamefont {Yang}}, \bibinfo {author} {\bibfnamefont {H.}~\bibnamefont {Zhou}}, \bibinfo {author} {\bibfnamefont {Z.-J.}\ \bibnamefont {Cheng}}, \bibinfo {author} {\bibfnamefont {Y.}~\bibnamefont {Huang}}, \bibinfo {author} {\bibfnamefont {Z.}~\bibnamefont {Qu}}, \bibinfo {author} {\bibfnamefont {F.}~\bibnamefont {Wang}}, \bibinfo {author} {\bibfnamefont {M.~Z.}\ \bibnamefont {Hasan}},\ and\ \bibinfo {author} {\bibfnamefont {S.}~\bibnamefont {Jia}},\ }\bibfield  {title} {\bibinfo {title} {Rare earth engineering in \text{$R{\mathrm{Mn}}_{6}{\mathrm{Sn}}_{6}$ ($R=\text{Gd}\text{\ensuremath{-}}\text{Tm}$, Lu)} topological kagome magnets},\ }\href {https://doi.org/10.1103/PhysRevLett.126.246602} {\bibfield  {journal} {\bibinfo  {journal} {Phys. Rev. Lett.}\ }\textbf {\bibinfo
  {volume} {126}},\ \bibinfo {pages} {246602} (\bibinfo {year} {2021})}\BibitemShut {NoStop}%
\bibitem [{\citenamefont {Fredrickson}\ \emph {et~al.}(2008)\citenamefont {Fredrickson}, \citenamefont {Lidin}, \citenamefont {Venturini}, \citenamefont {Malaman},\ and\ \citenamefont {Christensen}}]{fredrickson2008origins}%
  \BibitemOpen
  \bibfield  {author} {\bibinfo {author} {\bibfnamefont {D.~C.}\ \bibnamefont {Fredrickson}}, \bibinfo {author} {\bibfnamefont {S.}~\bibnamefont {Lidin}}, \bibinfo {author} {\bibfnamefont {G.}~\bibnamefont {Venturini}}, \bibinfo {author} {\bibfnamefont {B.}~\bibnamefont {Malaman}},\ and\ \bibinfo {author} {\bibfnamefont {J.}~\bibnamefont {Christensen}},\ }\bibfield  {title} {\bibinfo {title} {{Origins of superstructure ordering and incommensurability in stuffed CoSn-Type phases}},\ }\href@noop {} {\bibfield  {journal} {\bibinfo  {journal} {Journal of the American Chemical Society}\ }\textbf {\bibinfo {volume} {130}},\ \bibinfo {pages} {8195} (\bibinfo {year} {2008})}\BibitemShut {NoStop}%
\bibitem [{\citenamefont {El~Idrissi}\ \emph {et~al.}(1991)\citenamefont {El~Idrissi}, \citenamefont {Venturini},\ and\ \citenamefont {Malaman}}]{el1991crystal}%
  \BibitemOpen
  \bibfield  {author} {\bibinfo {author} {\bibfnamefont {B.~C.}\ \bibnamefont {El~Idrissi}}, \bibinfo {author} {\bibfnamefont {G.}~\bibnamefont {Venturini}},\ and\ \bibinfo {author} {\bibfnamefont {B.}~\bibnamefont {Malaman}},\ }\bibfield  {title} {\bibinfo {title} {Crystal structures of rfe6sn6 (r= sc, y, gd, tm, lu) rare-earth iron stannides},\ }\href@noop {} {\bibfield  {journal} {\bibinfo  {journal} {Materials Research Bulletin}\ }\textbf {\bibinfo {volume} {26}},\ \bibinfo {pages} {1331} (\bibinfo {year} {1991})}\BibitemShut {NoStop}%
\bibitem [{\citenamefont {Ortiz}\ \emph {et~al.}(2023{\natexlab{a}})\citenamefont {Ortiz}, \citenamefont {Pokharel}, \citenamefont {Gundayao}, \citenamefont {Li}, \citenamefont {Kaboudvand}, \citenamefont {Kautzsch}, \citenamefont {Sarker}, \citenamefont {Ruff}, \citenamefont {Hogan}, \citenamefont {Alvarado} \emph {et~al.}}]{ortiz2023ybv}%
  \BibitemOpen
  \bibfield  {author} {\bibinfo {author} {\bibfnamefont {B.~R.}\ \bibnamefont {Ortiz}}, \bibinfo {author} {\bibfnamefont {G.}~\bibnamefont {Pokharel}}, \bibinfo {author} {\bibfnamefont {M.}~\bibnamefont {Gundayao}}, \bibinfo {author} {\bibfnamefont {H.}~\bibnamefont {Li}}, \bibinfo {author} {\bibfnamefont {F.}~\bibnamefont {Kaboudvand}}, \bibinfo {author} {\bibfnamefont {L.}~\bibnamefont {Kautzsch}}, \bibinfo {author} {\bibfnamefont {S.}~\bibnamefont {Sarker}}, \bibinfo {author} {\bibfnamefont {J.~P.}\ \bibnamefont {Ruff}}, \bibinfo {author} {\bibfnamefont {T.}~\bibnamefont {Hogan}}, \bibinfo {author} {\bibfnamefont {S.~J.~G.}\ \bibnamefont {Alvarado}}, \emph {et~al.},\ }\bibfield  {title} {\bibinfo {title} {{YbV$_3$Sb$_4$ and EuV$_3$Sb$_4$ vanadium-based kagome metals with Yb$^{2+}$ and Eu$^{2+}$ zigzag chains}},\ }\href@noop {} {\bibfield  {journal} {\bibinfo  {journal} {Phys. Rev. Mater.}\ }\textbf {\bibinfo {volume} {7}},\ \bibinfo {pages} {064201} (\bibinfo {year} {2023}{\natexlab{a}})}\BibitemShut
  {NoStop}%
\bibitem [{\citenamefont {Ortiz}\ \emph {et~al.}(2023{\natexlab{b}})\citenamefont {Ortiz}, \citenamefont {Miao}, \citenamefont {Parker}, \citenamefont {Yang}, \citenamefont {Samolyuk}, \citenamefont {Clements}, \citenamefont {Rajapitamahuni}, \citenamefont {Yilmaz}, \citenamefont {Vescovo}, \citenamefont {Yan} \emph {et~al.}}]{ortiz2023evolution}%
  \BibitemOpen
  \bibfield  {author} {\bibinfo {author} {\bibfnamefont {B.~R.}\ \bibnamefont {Ortiz}}, \bibinfo {author} {\bibfnamefont {H.}~\bibnamefont {Miao}}, \bibinfo {author} {\bibfnamefont {D.~S.}\ \bibnamefont {Parker}}, \bibinfo {author} {\bibfnamefont {F.}~\bibnamefont {Yang}}, \bibinfo {author} {\bibfnamefont {G.~D.}\ \bibnamefont {Samolyuk}}, \bibinfo {author} {\bibfnamefont {E.~M.}\ \bibnamefont {Clements}}, \bibinfo {author} {\bibfnamefont {A.}~\bibnamefont {Rajapitamahuni}}, \bibinfo {author} {\bibfnamefont {T.}~\bibnamefont {Yilmaz}}, \bibinfo {author} {\bibfnamefont {E.}~\bibnamefont {Vescovo}}, \bibinfo {author} {\bibfnamefont {J.}~\bibnamefont {Yan}}, \emph {et~al.},\ }\bibfield  {title} {\bibinfo {title} {{Evolution of Highly Anisotropic Magnetism in the Titanium-Based Kagome Metals LnTi$_3$Bi$_4$ (Ln: La{\textperiodcentered}{\textperiodcentered}{\textperiodcentered} Gd$^{3+}$, Eu$^{2+}$, Yb$^{2+}$)}},\ }\href@noop {} {\bibfield  {journal} {\bibinfo  {journal} {Chemistry of Materials}\ }\textbf {\bibinfo
  {volume} {35}},\ \bibinfo {pages} {9756} (\bibinfo {year} {2023}{\natexlab{b}})}\BibitemShut {NoStop}%
\bibitem [{\citenamefont {Ortiz}\ \emph {et~al.}(2019)\citenamefont {Ortiz}, \citenamefont {Gomes}, \citenamefont {Morey}, \citenamefont {Winiarski}, \citenamefont {Bordelon}, \citenamefont {Mangum}, \citenamefont {Oswald}, \citenamefont {Rodriguez-Rivera}, \citenamefont {Neilson}, \citenamefont {Wilson} \emph {et~al.}}]{ortiz2019new}%
  \BibitemOpen
  \bibfield  {author} {\bibinfo {author} {\bibfnamefont {B.~R.}\ \bibnamefont {Ortiz}}, \bibinfo {author} {\bibfnamefont {L.~C.}\ \bibnamefont {Gomes}}, \bibinfo {author} {\bibfnamefont {J.~R.}\ \bibnamefont {Morey}}, \bibinfo {author} {\bibfnamefont {M.}~\bibnamefont {Winiarski}}, \bibinfo {author} {\bibfnamefont {M.}~\bibnamefont {Bordelon}}, \bibinfo {author} {\bibfnamefont {J.~S.}\ \bibnamefont {Mangum}}, \bibinfo {author} {\bibfnamefont {I.~W.}\ \bibnamefont {Oswald}}, \bibinfo {author} {\bibfnamefont {J.~A.}\ \bibnamefont {Rodriguez-Rivera}}, \bibinfo {author} {\bibfnamefont {J.~R.}\ \bibnamefont {Neilson}}, \bibinfo {author} {\bibfnamefont {S.~D.}\ \bibnamefont {Wilson}}, \emph {et~al.},\ }\bibfield  {title} {\bibinfo {title} {{New kagome prototype materials: discovery of KV$_3$Sb$_5$, RbV$_3$Sb$_5$, and CsV$_3$Sb$_5$}},\ }\href@noop {} {\bibfield  {journal} {\bibinfo  {journal} {Phys. Rev. Materials}\ }\textbf {\bibinfo {volume} {3}},\ \bibinfo {pages} {094407} (\bibinfo {year} {2019})}\BibitemShut
  {NoStop}%
\bibitem [{\citenamefont {Ovchinnikov}\ and\ \citenamefont {Bobev}(2018)}]{ovchinnikov2018synthesis}%
  \BibitemOpen
  \bibfield  {author} {\bibinfo {author} {\bibfnamefont {A.}~\bibnamefont {Ovchinnikov}}\ and\ \bibinfo {author} {\bibfnamefont {S.}~\bibnamefont {Bobev}},\ }\bibfield  {title} {\bibinfo {title} {{Synthesis, Crystal and Electronic Structure of the Titanium Bismuthides Sr$_5$Ti$_{12}$Bi$_{19+x}$, Ba$_5$Ti$_{12}$Bi$_{19+x}$, and Sr$_{5-\delta}$Eu$_\delta$Ti$_{12}$Bi$_{19+x}$ (x=0.5--1.0; $\delta$=2.4, 4.0)}},\ }\href@noop {} {\bibfield  {journal} {\bibinfo  {journal} {Eur. J. Inorg. Chem.}\ }\textbf {\bibinfo {volume} {2018}},\ \bibinfo {pages} {1266} (\bibinfo {year} {2018})}\BibitemShut {NoStop}%
\bibitem [{\citenamefont {Ovchinnikov}\ and\ \citenamefont {Bobev}(2019)}]{ovchinnikov2019bismuth}%
  \BibitemOpen
  \bibfield  {author} {\bibinfo {author} {\bibfnamefont {A.}~\bibnamefont {Ovchinnikov}}\ and\ \bibinfo {author} {\bibfnamefont {S.}~\bibnamefont {Bobev}},\ }\bibfield  {title} {\bibinfo {title} {Bismuth as a reactive solvent in the synthesis of multicomponent transition-metal-bearing bismuthides},\ }\href@noop {} {\bibfield  {journal} {\bibinfo  {journal} {Inorg. Chem.}\ }\textbf {\bibinfo {volume} {59}},\ \bibinfo {pages} {3459} (\bibinfo {year} {2019})}\BibitemShut {NoStop}%
\bibitem [{\citenamefont {Motoyama}\ \emph {et~al.}(2018)\citenamefont {Motoyama}, \citenamefont {Sezaki}, \citenamefont {Gouchi}, \citenamefont {Miyoshi}, \citenamefont {Nishigori}, \citenamefont {Mutou}, \citenamefont {Fujiwara},\ and\ \citenamefont {Uwatoko}}]{motoyama2018magnetic}%
  \BibitemOpen
  \bibfield  {author} {\bibinfo {author} {\bibfnamefont {G.}~\bibnamefont {Motoyama}}, \bibinfo {author} {\bibfnamefont {M.}~\bibnamefont {Sezaki}}, \bibinfo {author} {\bibfnamefont {J.}~\bibnamefont {Gouchi}}, \bibinfo {author} {\bibfnamefont {K.}~\bibnamefont {Miyoshi}}, \bibinfo {author} {\bibfnamefont {S.}~\bibnamefont {Nishigori}}, \bibinfo {author} {\bibfnamefont {T.}~\bibnamefont {Mutou}}, \bibinfo {author} {\bibfnamefont {K.}~\bibnamefont {Fujiwara}},\ and\ \bibinfo {author} {\bibfnamefont {Y.}~\bibnamefont {Uwatoko}},\ }\bibfield  {title} {\bibinfo {title} {{Magnetic properties of new antiferromagnetic heavy-fermion compounds, Ce$_3$TiBi$_5$ and CeTi$_3$Bi$_4$}},\ }\href@noop {} {\bibfield  {journal} {\bibinfo  {journal} {Physica B Condens.}\ }\textbf {\bibinfo {volume} {536}},\ \bibinfo {pages} {142} (\bibinfo {year} {2018})}\BibitemShut {NoStop}%
\bibitem [{\citenamefont {Chen}\ \emph {et~al.}(2024)\citenamefont {Chen}, \citenamefont {Zhou}, \citenamefont {Zhang}, \citenamefont {Ji}, \citenamefont {Liao}, \citenamefont {Ji}, \citenamefont {Li}, \citenamefont {Guo}, \citenamefont {Shen}, \citenamefont {Yu} \emph {et~al.}}]{chen2023134}%
  \BibitemOpen
  \bibfield  {author} {\bibinfo {author} {\bibfnamefont {L.}~\bibnamefont {Chen}}, \bibinfo {author} {\bibfnamefont {Y.}~\bibnamefont {Zhou}}, \bibinfo {author} {\bibfnamefont {H.}~\bibnamefont {Zhang}}, \bibinfo {author} {\bibfnamefont {X.}~\bibnamefont {Ji}}, \bibinfo {author} {\bibfnamefont {K.}~\bibnamefont {Liao}}, \bibinfo {author} {\bibfnamefont {Y.}~\bibnamefont {Ji}}, \bibinfo {author} {\bibfnamefont {Y.}~\bibnamefont {Li}}, \bibinfo {author} {\bibfnamefont {Z.}~\bibnamefont {Guo}}, \bibinfo {author} {\bibfnamefont {X.}~\bibnamefont {Shen}}, \bibinfo {author} {\bibfnamefont {R.}~\bibnamefont {Yu}}, \emph {et~al.},\ }\bibfield  {title} {\bibinfo {title} {Tunable magnetism in titanium-based kagome metals by rare-earth engineering and high pressure},\ }\href@noop {} {\bibfield  {journal} {\bibinfo  {journal} {Communications Materials}\ }\textbf {\bibinfo {volume} {5}},\ \bibinfo {pages} {73} (\bibinfo {year} {2024})}\BibitemShut {NoStop}%
\bibitem [{\citenamefont {Guo}\ \emph {et~al.}({\natexlab{a}})\citenamefont {Guo}, \citenamefont {Zhou}, \citenamefont {Ding}, \citenamefont {Qu}, \citenamefont {Liu}, \citenamefont {Du}, \citenamefont {Zhang}, \citenamefont {Li}, \citenamefont {Zhang}, \citenamefont {Zhou}, \citenamefont {Qi}, \citenamefont {Guo}, \citenamefont {Wang}, \citenamefont {Fei}, \citenamefont {Huang}, \citenamefont {Qian}, \citenamefont {Shen}, \citenamefont {Weng},\ and\ \citenamefont {Song}}]{guo2023134}%
  \BibitemOpen
  \bibfield  {author} {\bibinfo {author} {\bibfnamefont {J.}~\bibnamefont {Guo}}, \bibinfo {author} {\bibfnamefont {L.}~\bibnamefont {Zhou}}, \bibinfo {author} {\bibfnamefont {J.}~\bibnamefont {Ding}}, \bibinfo {author} {\bibfnamefont {G.}~\bibnamefont {Qu}}, \bibinfo {author} {\bibfnamefont {Z.}~\bibnamefont {Liu}}, \bibinfo {author} {\bibfnamefont {Y.}~\bibnamefont {Du}}, \bibinfo {author} {\bibfnamefont {H.}~\bibnamefont {Zhang}}, \bibinfo {author} {\bibfnamefont {J.}~\bibnamefont {Li}}, \bibinfo {author} {\bibfnamefont {Y.}~\bibnamefont {Zhang}}, \bibinfo {author} {\bibfnamefont {F.}~\bibnamefont {Zhou}}, \bibinfo {author} {\bibfnamefont {W.}~\bibnamefont {Qi}}, \bibinfo {author} {\bibfnamefont {F.}~\bibnamefont {Guo}}, \bibinfo {author} {\bibfnamefont {T.}~\bibnamefont {Wang}}, \bibinfo {author} {\bibfnamefont {F.}~\bibnamefont {Fei}}, \bibinfo {author} {\bibfnamefont {Y.}~\bibnamefont {Huang}}, \bibinfo {author} {\bibfnamefont {T.}~\bibnamefont {Qian}}, \bibinfo {author} {\bibfnamefont {D.}~\bibnamefont
  {Shen}}, \bibinfo {author} {\bibfnamefont {H.}~\bibnamefont {Weng}},\ and\ \bibinfo {author} {\bibfnamefont {F.}~\bibnamefont {Song}},\ }\bibfield  {title} {\bibinfo {title} {{Magnetic kagome materials RETi$_3$Bi$_4$ family with weak interlayer interactions}},\ }\href@noop {} {\bibfield  {journal} {\bibinfo  {journal} {\textbf{2023}, arXiv:2308.14509v1 [cond-mat.mtrl-sci]. arXiv.org e-Print archive. https://arxiv.org/pdf/2308.14509.pdf (Accessed 10-24-2023)}\ } ({\natexlab{a}})}\BibitemShut {NoStop}%
\bibitem [{\citenamefont {Jovanovic}\ and\ \citenamefont {Schoop}(2022)}]{jovanovic2022simple}%
  \BibitemOpen
  \bibfield  {author} {\bibinfo {author} {\bibfnamefont {M.}~\bibnamefont {Jovanovic}}\ and\ \bibinfo {author} {\bibfnamefont {L.~M.}\ \bibnamefont {Schoop}},\ }\bibfield  {title} {\bibinfo {title} {{Simple chemical rules for predicting band structures of kagome materials}},\ }\href@noop {} {\bibfield  {journal} {\bibinfo  {journal} {Journal of the American Chemical Society}\ }\textbf {\bibinfo {volume} {144}},\ \bibinfo {pages} {10978} (\bibinfo {year} {2022})}\BibitemShut {NoStop}%
\bibitem [{\citenamefont {Zheng}\ \emph {et~al.}(2024)\citenamefont {Zheng}, \citenamefont {Chen}, \citenamefont {Ji}, \citenamefont {Zhou}, \citenamefont {Qu}, \citenamefont {Hu}, \citenamefont {Huang}, \citenamefont {Weng}, \citenamefont {Qian},\ and\ \citenamefont {Wang}}]{chen2023sm134}%
  \BibitemOpen
  \bibfield  {author} {\bibinfo {author} {\bibfnamefont {Z.}~\bibnamefont {Zheng}}, \bibinfo {author} {\bibfnamefont {L.}~\bibnamefont {Chen}}, \bibinfo {author} {\bibfnamefont {X.}~\bibnamefont {Ji}}, \bibinfo {author} {\bibfnamefont {Y.}~\bibnamefont {Zhou}}, \bibinfo {author} {\bibfnamefont {G.}~\bibnamefont {Qu}}, \bibinfo {author} {\bibfnamefont {M.}~\bibnamefont {Hu}}, \bibinfo {author} {\bibfnamefont {Y.}~\bibnamefont {Huang}}, \bibinfo {author} {\bibfnamefont {H.}~\bibnamefont {Weng}}, \bibinfo {author} {\bibfnamefont {T.}~\bibnamefont {Qian}},\ and\ \bibinfo {author} {\bibfnamefont {G.}~\bibnamefont {Wang}},\ }\bibfield  {title} {\bibinfo {title} {{Anisotropic magnetism and band evolution induced by ferromagnetic phase transition in titanium-based kagome ferromagnet SmTi$_3$Bi$_4$}},\ }\href@noop {} {\bibfield  {journal} {\bibinfo  {journal} {Science China Physics, Mechanics \& Astronomy}\ }\textbf {\bibinfo {volume} {67}},\ \bibinfo {pages} {267411} (\bibinfo {year} {2024})}\BibitemShut {NoStop}%
\bibitem [{\citenamefont {Hu}\ \emph {et~al.}()\citenamefont {Hu}, \citenamefont {Le}, \citenamefont {Chen}, \citenamefont {Deng}, \citenamefont {Zhou}, \citenamefont {Plumb}, \citenamefont {Radovic}, \citenamefont {Thomale}, \citenamefont {Schnyder}, \citenamefont {Yin} \emph {et~al.}}]{hu2023magnetic}%
  \BibitemOpen
  \bibfield  {author} {\bibinfo {author} {\bibfnamefont {Y.}~\bibnamefont {Hu}}, \bibinfo {author} {\bibfnamefont {C.}~\bibnamefont {Le}}, \bibinfo {author} {\bibfnamefont {L.}~\bibnamefont {Chen}}, \bibinfo {author} {\bibfnamefont {H.}~\bibnamefont {Deng}}, \bibinfo {author} {\bibfnamefont {Y.}~\bibnamefont {Zhou}}, \bibinfo {author} {\bibfnamefont {N.~C.}\ \bibnamefont {Plumb}}, \bibinfo {author} {\bibfnamefont {M.}~\bibnamefont {Radovic}}, \bibinfo {author} {\bibfnamefont {R.}~\bibnamefont {Thomale}}, \bibinfo {author} {\bibfnamefont {A.~P.}\ \bibnamefont {Schnyder}}, \bibinfo {author} {\bibfnamefont {J.-X.}\ \bibnamefont {Yin}}, \emph {et~al.},\ }\bibfield  {title} {\bibinfo {title} {Magnetic-coupled electronic landscape in bilayer-distorted titanium-based kagome metals},\ }\href@noop {} {\bibinfo  {journal} {\textbf{2023}, arXiv:2311.07747v1 [cond-mat.mtrl-sci]. arXiv.org e-Print archive. https://arxiv.org/pdf/2311.07747v1 (Accessed 6-25-2024)}\ }\BibitemShut {NoStop}%
\bibitem [{\citenamefont {Mondal}\ \emph {et~al.}()\citenamefont {Mondal}, \citenamefont {Sakhya}, \citenamefont {Sprague}, \citenamefont {Ortiz}, \citenamefont {Matzelle}, \citenamefont {Ghosh}, \citenamefont {Valadez}, \citenamefont {Elius}, \citenamefont {Bansil},\ and\ \citenamefont {Neupane}}]{mondal2023observation}%
  \BibitemOpen
\bibfield  {journal} {  }\bibfield  {author} {\bibinfo {author} {\bibfnamefont {M.~I.}\ \bibnamefont {Mondal}}, \bibinfo {author} {\bibfnamefont {A.~P.}\ \bibnamefont {Sakhya}}, \bibinfo {author} {\bibfnamefont {M.}~\bibnamefont {Sprague}}, \bibinfo {author} {\bibfnamefont {B.~R.}\ \bibnamefont {Ortiz}}, \bibinfo {author} {\bibfnamefont {M.}~\bibnamefont {Matzelle}}, \bibinfo {author} {\bibfnamefont {B.}~\bibnamefont {Ghosh}}, \bibinfo {author} {\bibfnamefont {N.}~\bibnamefont {Valadez}}, \bibinfo {author} {\bibfnamefont {I.~B.}\ \bibnamefont {Elius}}, \bibinfo {author} {\bibfnamefont {A.}~\bibnamefont {Bansil}},\ and\ \bibinfo {author} {\bibfnamefont {M.}~\bibnamefont {Neupane}},\ }\bibfield  {title} {\bibinfo {title} {{Observation of multiple van Hove singularities and correlated electronic states in a new topological ferromagnetic kagome metal NdTi$_3$Bi$_4$}},\ }\href@noop {} {\bibinfo  {journal} {\textbf{2023}, arXiv:2311.11488v1 [cond-mat.mtrl-sci]. arXiv.org e-Print archive.
  https://arxiv.org/pdf/2311.11488v1 (Accessed 6-25-2024)}\ }\BibitemShut {NoStop}%
\bibitem [{\citenamefont {Jiang}\ \emph {et~al.}()\citenamefont {Jiang}, \citenamefont {Li}, \citenamefont {Yuan}, \citenamefont {Liu}, \citenamefont {Cao}, \citenamefont {Cho}, \citenamefont {Shu}, \citenamefont {Yang}, \citenamefont {Ding}, \citenamefont {Li} \emph {et~al.}}]{jiang2023direct}%
  \BibitemOpen
\bibfield  {journal} {  }\bibfield  {author} {\bibinfo {author} {\bibfnamefont {Z.}~\bibnamefont {Jiang}}, \bibinfo {author} {\bibfnamefont {T.}~\bibnamefont {Li}}, \bibinfo {author} {\bibfnamefont {J.}~\bibnamefont {Yuan}}, \bibinfo {author} {\bibfnamefont {Z.}~\bibnamefont {Liu}}, \bibinfo {author} {\bibfnamefont {Z.}~\bibnamefont {Cao}}, \bibinfo {author} {\bibfnamefont {S.}~\bibnamefont {Cho}}, \bibinfo {author} {\bibfnamefont {M.}~\bibnamefont {Shu}}, \bibinfo {author} {\bibfnamefont {Y.}~\bibnamefont {Yang}}, \bibinfo {author} {\bibfnamefont {J.}~\bibnamefont {Ding}}, \bibinfo {author} {\bibfnamefont {Z.}~\bibnamefont {Li}}, \emph {et~al.},\ }\bibfield  {title} {\bibinfo {title} {Direct observation of topological surface states in the layered kagome lattice with broken time-reversal symmetry},\ }\href@noop {} {\bibinfo  {journal} {\textbf{2023}, arXiv:2309.01579v1 [cond-mat.str-el]. arXiv.org e-Print archive. https://arxiv.org/pdf/2309.01579v1 (Accessed 6-25-2024)}\ }\BibitemShut {NoStop}%
\bibitem [{\citenamefont {Sakhya}\ \emph {et~al.}()\citenamefont {Sakhya}, \citenamefont {Ortiz}, \citenamefont {Ghosh}, \citenamefont {Sprague}, \citenamefont {Mondal}, \citenamefont {Matzelle}, \citenamefont {Elius}, \citenamefont {Valadez}, \citenamefont {Mandrus}, \citenamefont {Bansil} \emph {et~al.}}]{sakhya2023observation}%
  \BibitemOpen
\bibfield  {journal} {  }\bibfield  {author} {\bibinfo {author} {\bibfnamefont {A.~P.}\ \bibnamefont {Sakhya}}, \bibinfo {author} {\bibfnamefont {B.~R.}\ \bibnamefont {Ortiz}}, \bibinfo {author} {\bibfnamefont {B.}~\bibnamefont {Ghosh}}, \bibinfo {author} {\bibfnamefont {M.}~\bibnamefont {Sprague}}, \bibinfo {author} {\bibfnamefont {M.~I.}\ \bibnamefont {Mondal}}, \bibinfo {author} {\bibfnamefont {M.}~\bibnamefont {Matzelle}}, \bibinfo {author} {\bibfnamefont {I.~B.}\ \bibnamefont {Elius}}, \bibinfo {author} {\bibfnamefont {N.}~\bibnamefont {Valadez}}, \bibinfo {author} {\bibfnamefont {D.~G.}\ \bibnamefont {Mandrus}}, \bibinfo {author} {\bibfnamefont {A.}~\bibnamefont {Bansil}}, \emph {et~al.},\ }\bibfield  {title} {\bibinfo {title} {{Observation of multiple flat bands and topological Dirac states in a new titanium based slightly distorted kagome metal YbTi$_3$Bi$_4$}},\ }\href@noop {} {\bibinfo  {journal} {\textbf{2023}, arXiv:2309.01176v1 [cond-mat.mes-hall]. arXiv.org e-Print archive.
  https://arxiv.org/pdf/2309.01176v1 (Accessed 6-25-2024)}\ }\BibitemShut {NoStop}%
\bibitem [{\citenamefont {Canfield}\ \emph {et~al.}(2016)\citenamefont {Canfield}, \citenamefont {Kong}, \citenamefont {Kaluarachchi},\ and\ \citenamefont {Jo}}]{canfield2016use}%
  \BibitemOpen
\bibfield  {journal} {  }\bibfield  {author} {\bibinfo {author} {\bibfnamefont {P.~C.}\ \bibnamefont {Canfield}}, \bibinfo {author} {\bibfnamefont {T.}~\bibnamefont {Kong}}, \bibinfo {author} {\bibfnamefont {U.~S.}\ \bibnamefont {Kaluarachchi}},\ and\ \bibinfo {author} {\bibfnamefont {N.~H.}\ \bibnamefont {Jo}},\ }\bibfield  {title} {\bibinfo {title} {Use of frit-disc crucibles for routine and exploratory solution growth of single crystalline samples},\ }\href@noop {} {\bibfield  {journal} {\bibinfo  {journal} {Philosophical magazine}\ }\textbf {\bibinfo {volume} {96}},\ \bibinfo {pages} {84} (\bibinfo {year} {2016})}\BibitemShut {NoStop}%
\bibitem [{\citenamefont {Taddei}\ \emph {et~al.}(2023)\citenamefont {Taddei}, \citenamefont {Garlea}, \citenamefont {Samarakoon}, \citenamefont {Sanjeewa}, \citenamefont {Xing}, \citenamefont {Heitmann}, \citenamefont {dela Cruz}, \citenamefont {Sefat},\ and\ \citenamefont {Parker}}]{taddei2023zigzag}%
  \BibitemOpen
  \bibfield  {author} {\bibinfo {author} {\bibfnamefont {K.}~\bibnamefont {Taddei}}, \bibinfo {author} {\bibfnamefont {V.}~\bibnamefont {Garlea}}, \bibinfo {author} {\bibfnamefont {A.}~\bibnamefont {Samarakoon}}, \bibinfo {author} {\bibfnamefont {L.}~\bibnamefont {Sanjeewa}}, \bibinfo {author} {\bibfnamefont {J.}~\bibnamefont {Xing}}, \bibinfo {author} {\bibfnamefont {T.}~\bibnamefont {Heitmann}}, \bibinfo {author} {\bibfnamefont {C.}~\bibnamefont {dela Cruz}}, \bibinfo {author} {\bibfnamefont {A.}~\bibnamefont {Sefat}},\ and\ \bibinfo {author} {\bibfnamefont {D.}~\bibnamefont {Parker}},\ }\bibfield  {title} {\bibinfo {title} {{Zigzag magnetic order and possible Kitaev interactions in the spin-1 honeycomb lattice KNiAsO$_4$}},\ }\href@noop {} {\bibfield  {journal} {\bibinfo  {journal} {Physical Review Research}\ }\textbf {\bibinfo {volume} {5}},\ \bibinfo {pages} {013022} (\bibinfo {year} {2023})}\BibitemShut {NoStop}%
\bibitem [{\citenamefont {Sales}\ \emph {et~al.}(2022)\citenamefont {Sales}, \citenamefont {Meier}, \citenamefont {Parker}, \citenamefont {Yin}, \citenamefont {Yan}, \citenamefont {May}, \citenamefont {Calder}, \citenamefont {Aczel}, \citenamefont {Zhang}, \citenamefont {Li} \emph {et~al.}}]{sales2022chemical}%
  \BibitemOpen
  \bibfield  {author} {\bibinfo {author} {\bibfnamefont {B.~C.}\ \bibnamefont {Sales}}, \bibinfo {author} {\bibfnamefont {W.~R.}\ \bibnamefont {Meier}}, \bibinfo {author} {\bibfnamefont {D.~S.}\ \bibnamefont {Parker}}, \bibinfo {author} {\bibfnamefont {L.}~\bibnamefont {Yin}}, \bibinfo {author} {\bibfnamefont {J.}~\bibnamefont {Yan}}, \bibinfo {author} {\bibfnamefont {A.~F.}\ \bibnamefont {May}}, \bibinfo {author} {\bibfnamefont {S.}~\bibnamefont {Calder}}, \bibinfo {author} {\bibfnamefont {A.~A.}\ \bibnamefont {Aczel}}, \bibinfo {author} {\bibfnamefont {Q.}~\bibnamefont {Zhang}}, \bibinfo {author} {\bibfnamefont {H.}~\bibnamefont {Li}}, \emph {et~al.},\ }\bibfield  {title} {\bibinfo {title} {{Chemical Control of Magnetism in the Kagome Metal CoSn$_{1--x}$In$_x$: Magnetic Order from Nonmagnetic Substitutions}},\ }\href@noop {} {\bibfield  {journal} {\bibinfo  {journal} {Chemistry of Materials}\ }\textbf {\bibinfo {volume} {34}},\ \bibinfo {pages} {7069} (\bibinfo {year} {2022})}\BibitemShut {NoStop}%
\bibitem [{\citenamefont {Sala}\ \emph {et~al.}(2022)\citenamefont {Sala}, \citenamefont {Lin}, \citenamefont {Samarakoon}, \citenamefont {Parker}, \citenamefont {May},\ and\ \citenamefont {Stone}}]{sala2022ferrimagnetic}%
  \BibitemOpen
  \bibfield  {author} {\bibinfo {author} {\bibfnamefont {G.}~\bibnamefont {Sala}}, \bibinfo {author} {\bibfnamefont {J.~Y.}\ \bibnamefont {Lin}}, \bibinfo {author} {\bibfnamefont {A.~M.}\ \bibnamefont {Samarakoon}}, \bibinfo {author} {\bibfnamefont {D.~S.}\ \bibnamefont {Parker}}, \bibinfo {author} {\bibfnamefont {A.~F.}\ \bibnamefont {May}},\ and\ \bibinfo {author} {\bibfnamefont {M.~B.}\ \bibnamefont {Stone}},\ }\bibfield  {title} {\bibinfo {title} {{Ferrimagnetic spin waves in honeycomb and triangular layers of Mn$_3$S$_2$Te$_6$}},\ }\href@noop {} {\bibfield  {journal} {\bibinfo  {journal} {Physical Review B}\ }\textbf {\bibinfo {volume} {105}},\ \bibinfo {pages} {214405} (\bibinfo {year} {2022})}\BibitemShut {NoStop}%
\bibitem [{\citenamefont {Hobbis}\ \emph {et~al.}(2019)\citenamefont {Hobbis}, \citenamefont {Hermann}, \citenamefont {Wang}, \citenamefont {Parker}, \citenamefont {Pandey}, \citenamefont {Martin}, \citenamefont {Page},\ and\ \citenamefont {Nolas}}]{hobbis2019structural}%
  \BibitemOpen
  \bibfield  {author} {\bibinfo {author} {\bibfnamefont {D.}~\bibnamefont {Hobbis}}, \bibinfo {author} {\bibfnamefont {R.~P.}\ \bibnamefont {Hermann}}, \bibinfo {author} {\bibfnamefont {H.}~\bibnamefont {Wang}}, \bibinfo {author} {\bibfnamefont {D.~S.}\ \bibnamefont {Parker}}, \bibinfo {author} {\bibfnamefont {T.}~\bibnamefont {Pandey}}, \bibinfo {author} {\bibfnamefont {J.}~\bibnamefont {Martin}}, \bibinfo {author} {\bibfnamefont {K.}~\bibnamefont {Page}},\ and\ \bibinfo {author} {\bibfnamefont {G.~S.}\ \bibnamefont {Nolas}},\ }\bibfield  {title} {\bibinfo {title} {{Structural, chemical, electrical, and thermal properties of n-type NbFeSb}},\ }\href@noop {} {\bibfield  {journal} {\bibinfo  {journal} {Inorganic chemistry}\ }\textbf {\bibinfo {volume} {58}},\ \bibinfo {pages} {1826} (\bibinfo {year} {2019})}\BibitemShut {NoStop}%
\bibitem [{\citenamefont {Pokharel}\ \emph {et~al.}(2018)\citenamefont {Pokharel}, \citenamefont {May}, \citenamefont {Parker}, \citenamefont {Calder}, \citenamefont {Ehlers}, \citenamefont {Huq}, \citenamefont {Kimber}, \citenamefont {Arachchige}, \citenamefont {Poudel}, \citenamefont {McGuire} \emph {et~al.}}]{pokharel2018negative}%
  \BibitemOpen
  \bibfield  {author} {\bibinfo {author} {\bibfnamefont {G.}~\bibnamefont {Pokharel}}, \bibinfo {author} {\bibfnamefont {A.}~\bibnamefont {May}}, \bibinfo {author} {\bibfnamefont {D.}~\bibnamefont {Parker}}, \bibinfo {author} {\bibfnamefont {S.}~\bibnamefont {Calder}}, \bibinfo {author} {\bibfnamefont {G.}~\bibnamefont {Ehlers}}, \bibinfo {author} {\bibfnamefont {A.}~\bibnamefont {Huq}}, \bibinfo {author} {\bibfnamefont {S.}~\bibnamefont {Kimber}}, \bibinfo {author} {\bibfnamefont {H.~S.}\ \bibnamefont {Arachchige}}, \bibinfo {author} {\bibfnamefont {L.}~\bibnamefont {Poudel}}, \bibinfo {author} {\bibfnamefont {M.}~\bibnamefont {McGuire}}, \emph {et~al.},\ }\bibfield  {title} {\bibinfo {title} {{Negative thermal expansion and magnetoelastic coupling in the breathing pyrochlore lattice material LiGaCr$_4$S$_8$}},\ }\href@noop {} {\bibfield  {journal} {\bibinfo  {journal} {Physical Review B}\ }\textbf {\bibinfo {volume} {97}},\ \bibinfo {pages} {134117} (\bibinfo {year} {2018})}\BibitemShut {NoStop}%
\bibitem [{\citenamefont {Richter}\ and\ \citenamefont {Jeitschko}(1997)}]{richter1997preparation}%
  \BibitemOpen
  \bibfield  {author} {\bibinfo {author} {\bibfnamefont {C.~G.}\ \bibnamefont {Richter}}\ and\ \bibinfo {author} {\bibfnamefont {W.}~\bibnamefont {Jeitschko}},\ }\bibfield  {title} {\bibinfo {title} {{Preparation and Crystal Structure of the Titanium and Hafnium Bismuthides Ti$_8$Bi$_9$ and Hf$_8$Bi$_9$}},\ }\href@noop {} {\bibfield  {journal} {\bibinfo  {journal} {Journal of Solid State Chemistry}\ }\textbf {\bibinfo {volume} {134}},\ \bibinfo {pages} {26} (\bibinfo {year} {1997})}\BibitemShut {NoStop}%
\bibitem [{\citenamefont {Vassilev}(2006)}]{vassilev2006contribution}%
  \BibitemOpen
  \bibfield  {author} {\bibinfo {author} {\bibfnamefont {G.}~\bibnamefont {Vassilev}},\ }\bibfield  {title} {\bibinfo {title} {{Contribution to the Ti-Bi phase diagram}},\ }\href@noop {} {\bibfield  {journal} {\bibinfo  {journal} {Crystal Research and Technology: Journal of Experimental and Industrial Crystallography}\ }\textbf {\bibinfo {volume} {41}},\ \bibinfo {pages} {349} (\bibinfo {year} {2006})}\BibitemShut {NoStop}%
\bibitem [{\citenamefont {Okamoto}(1998)}]{okamoto1998bi}%
  \BibitemOpen
  \bibfield  {author} {\bibinfo {author} {\bibfnamefont {H.}~\bibnamefont {Okamoto}},\ }\bibfield  {title} {\bibinfo {title} {{Bi-Tb (bismuth-terbium)}},\ }\href@noop {} {\bibfield  {journal} {\bibinfo  {journal} {Journal of phase equilibria}\ }\textbf {\bibinfo {volume} {19}},\ \bibinfo {pages} {489} (\bibinfo {year} {1998})}\BibitemShut {NoStop}%
\bibitem [{\citenamefont {Yamane}\ and\ \citenamefont {Hiraka}(2018)}]{yamane2018crystal}%
  \BibitemOpen
  \bibfield  {author} {\bibinfo {author} {\bibfnamefont {H.}~\bibnamefont {Yamane}}\ and\ \bibinfo {author} {\bibfnamefont {K.}~\bibnamefont {Hiraka}},\ }\bibfield  {title} {\bibinfo {title} {{Crystal structure of Ti$_8$Bi$_9$O$_{0.25}$ containing interstitial oxygen atoms}},\ }\href@noop {} {\bibfield  {journal} {\bibinfo  {journal} {Acta Crystallographica Section E: Crystallographic Communications}\ }\textbf {\bibinfo {volume} {74}},\ \bibinfo {pages} {1366} (\bibinfo {year} {2018})}\BibitemShut {NoStop}%
\bibitem [{\citenamefont {Cheng}\ \emph {et~al.}()\citenamefont {Cheng}, \citenamefont {Wang}, \citenamefont {Nie}, \citenamefont {Ying}, \citenamefont {Li}, \citenamefont {Li}, \citenamefont {Xu}, \citenamefont {Chen}, \citenamefont {Koban}, \citenamefont {Borrmann} \emph {et~al.}}]{cheng2024giant}%
  \BibitemOpen
  \bibfield  {author} {\bibinfo {author} {\bibfnamefont {E.}~\bibnamefont {Cheng}}, \bibinfo {author} {\bibfnamefont {K.}~\bibnamefont {Wang}}, \bibinfo {author} {\bibfnamefont {S.}~\bibnamefont {Nie}}, \bibinfo {author} {\bibfnamefont {T.}~\bibnamefont {Ying}}, \bibinfo {author} {\bibfnamefont {Z.}~\bibnamefont {Li}}, \bibinfo {author} {\bibfnamefont {Y.}~\bibnamefont {Li}}, \bibinfo {author} {\bibfnamefont {Y.}~\bibnamefont {Xu}}, \bibinfo {author} {\bibfnamefont {H.}~\bibnamefont {Chen}}, \bibinfo {author} {\bibfnamefont {R.}~\bibnamefont {Koban}}, \bibinfo {author} {\bibfnamefont {H.}~\bibnamefont {Borrmann}}, \emph {et~al.},\ }\bibfield  {title} {\bibinfo {title} {Giant anomalous hall effect and band folding in a kagome metal with mixed dimensionality},\ }\href@noop {} {\bibinfo  {journal} {\textbf{2024}, arXiv:2405.16831v1 [cond-mat.str-el]. arXiv.org e-Print archive. https://arxiv.org/pdf/2405.16831v1 (Accessed 6-25-2024)}\ }\BibitemShut {NoStop}%
\bibitem [{\citenamefont {Guo}\ \emph {et~al.}({\natexlab{b}})\citenamefont {Guo}, \citenamefont {Ma}, \citenamefont {Liu}, \citenamefont {Wu}, \citenamefont {Wang}, \citenamefont {Shi}, \citenamefont {Li},\ and\ \citenamefont {Jia}}]{guo20241}%
  \BibitemOpen
\bibfield  {journal} {  }\bibfield  {author} {\bibinfo {author} {\bibfnamefont {K.}~\bibnamefont {Guo}}, \bibinfo {author} {\bibfnamefont {Z.}~\bibnamefont {Ma}}, \bibinfo {author} {\bibfnamefont {H.}~\bibnamefont {Liu}}, \bibinfo {author} {\bibfnamefont {Z.}~\bibnamefont {Wu}}, \bibinfo {author} {\bibfnamefont {J.}~\bibnamefont {Wang}}, \bibinfo {author} {\bibfnamefont {Y.}~\bibnamefont {Shi}}, \bibinfo {author} {\bibfnamefont {Y.}~\bibnamefont {Li}},\ and\ \bibinfo {author} {\bibfnamefont {S.}~\bibnamefont {Jia}},\ }\bibfield  {title} {\bibinfo {title} {{1/3 and other magnetization plateaus in a quasi-one-dimensional Ising magnet TbTi$_3$Bi$_4$ with zigzag spin chain}},\ }\href@noop {} {\bibfield  {journal} {\bibinfo  {journal} {\textbf{2024}, arXiv:2405.09280v1 [cond-mat.str-el]. arXiv.org e-Print archive. https://arxiv.org/pdf/2405.09280v1 (Accessed 6-25-2024)}\ } ({\natexlab{b}})}\BibitemShut {NoStop}%
\bibitem [{\citenamefont {Heltemes}\ and\ \citenamefont {Swenson}(1961)}]{heltemes1961nuclear}%
  \BibitemOpen
  \bibfield  {author} {\bibinfo {author} {\bibfnamefont {E.}~\bibnamefont {Heltemes}}\ and\ \bibinfo {author} {\bibfnamefont {C.}~\bibnamefont {Swenson}},\ }\bibfield  {title} {\bibinfo {title} {Nuclear contribution to the heat capacity of terbium metal},\ }\href@noop {} {\bibfield  {journal} {\bibinfo  {journal} {The Journal of Chemical Physics}\ }\textbf {\bibinfo {volume} {35}},\ \bibinfo {pages} {1264} (\bibinfo {year} {1961})}\BibitemShut {NoStop}%
\bibitem [{\citenamefont {Jungwirth}\ \emph {et~al.}(2016)\citenamefont {Jungwirth}, \citenamefont {Marti}, \citenamefont {Wadley},\ and\ \citenamefont {Wunderlich}}]{jungwirth2016antiferromagnetic}%
  \BibitemOpen
  \bibfield  {author} {\bibinfo {author} {\bibfnamefont {T.}~\bibnamefont {Jungwirth}}, \bibinfo {author} {\bibfnamefont {X.}~\bibnamefont {Marti}}, \bibinfo {author} {\bibfnamefont {P.}~\bibnamefont {Wadley}},\ and\ \bibinfo {author} {\bibfnamefont {J.}~\bibnamefont {Wunderlich}},\ }\bibfield  {title} {\bibinfo {title} {Antiferromagnetic spintronics},\ }\href@noop {} {\bibfield  {journal} {\bibinfo  {journal} {Nature nanotechnology}\ }\textbf {\bibinfo {volume} {11}},\ \bibinfo {pages} {231} (\bibinfo {year} {2016})}\BibitemShut {NoStop}%
\bibitem [{\citenamefont {Baltz}\ \emph {et~al.}(2018)\citenamefont {Baltz}, \citenamefont {Manchon}, \citenamefont {Tsoi}, \citenamefont {Moriyama}, \citenamefont {Ono},\ and\ \citenamefont {Tserkovnyak}}]{baltz2018antiferromagnetic}%
  \BibitemOpen
  \bibfield  {author} {\bibinfo {author} {\bibfnamefont {V.}~\bibnamefont {Baltz}}, \bibinfo {author} {\bibfnamefont {A.}~\bibnamefont {Manchon}}, \bibinfo {author} {\bibfnamefont {M.}~\bibnamefont {Tsoi}}, \bibinfo {author} {\bibfnamefont {T.}~\bibnamefont {Moriyama}}, \bibinfo {author} {\bibfnamefont {T.}~\bibnamefont {Ono}},\ and\ \bibinfo {author} {\bibfnamefont {Y.}~\bibnamefont {Tserkovnyak}},\ }\bibfield  {title} {\bibinfo {title} {Antiferromagnetic spintronics},\ }\href@noop {} {\bibfield  {journal} {\bibinfo  {journal} {Reviews of Modern Physics}\ }\textbf {\bibinfo {volume} {90}},\ \bibinfo {pages} {015005} (\bibinfo {year} {2018})}\BibitemShut {NoStop}%
\bibitem [{\citenamefont {Jain}\ \emph {et~al.}(2013)\citenamefont {Jain}, \citenamefont {Ong}, \citenamefont {Hautier}, \citenamefont {Chen}, \citenamefont {Richards}, \citenamefont {Dacek}, \citenamefont {Cholia}, \citenamefont {Gunter}, \citenamefont {Skinner}, \citenamefont {Ceder},\ and\ \citenamefont {Persson}}]{Jain2013}%
  \BibitemOpen
  \bibfield  {author} {\bibinfo {author} {\bibfnamefont {A.}~\bibnamefont {Jain}}, \bibinfo {author} {\bibfnamefont {S.~P.}\ \bibnamefont {Ong}}, \bibinfo {author} {\bibfnamefont {G.}~\bibnamefont {Hautier}}, \bibinfo {author} {\bibfnamefont {W.}~\bibnamefont {Chen}}, \bibinfo {author} {\bibfnamefont {W.~D.}\ \bibnamefont {Richards}}, \bibinfo {author} {\bibfnamefont {S.}~\bibnamefont {Dacek}}, \bibinfo {author} {\bibfnamefont {S.}~\bibnamefont {Cholia}}, \bibinfo {author} {\bibfnamefont {D.}~\bibnamefont {Gunter}}, \bibinfo {author} {\bibfnamefont {D.}~\bibnamefont {Skinner}}, \bibinfo {author} {\bibfnamefont {G.}~\bibnamefont {Ceder}},\ and\ \bibinfo {author} {\bibfnamefont {K.~a.}\ \bibnamefont {Persson}},\ }\bibfield  {title} {\bibinfo {title} {{The Materials Project: A materials genome approach to accelerating materials innovation}},\ }\href {https://doi.org/10.1063/1.4812323} {\bibfield  {journal} {\bibinfo  {journal} {APL Materials}\ }\textbf {\bibinfo {volume} {1}},\ \bibinfo {pages} {011002}
  (\bibinfo {year} {2013})}\BibitemShut {NoStop}%
\bibitem [{\citenamefont {Setyawan}\ and\ \citenamefont {Curtarolo}(2010)}]{setyawan2010high}%
  \BibitemOpen
  \bibfield  {author} {\bibinfo {author} {\bibfnamefont {W.}~\bibnamefont {Setyawan}}\ and\ \bibinfo {author} {\bibfnamefont {S.}~\bibnamefont {Curtarolo}},\ }\bibfield  {title} {\bibinfo {title} {High-throughput electronic band structure calculations: Challenges and tools},\ }\href@noop {} {\bibfield  {journal} {\bibinfo  {journal} {Comput. Mater. Sci.}\ }\textbf {\bibinfo {volume} {49}},\ \bibinfo {pages} {299} (\bibinfo {year} {2010})}\BibitemShut {NoStop}%
\bibitem [{\citenamefont {Kleinke}(1998)}]{kleinke1998stabilization}%
  \BibitemOpen
  \bibfield  {author} {\bibinfo {author} {\bibfnamefont {H.}~\bibnamefont {Kleinke}},\ }\bibfield  {title} {\bibinfo {title} {{Stabilization of the New Antimonide Zr$_2$V$_6$Sb$_9$ by V--V and Sb--Sb Bonding}},\ }\href@noop {} {\bibfield  {journal} {\bibinfo  {journal} {European journal of inorganic chemistry}\ }\textbf {\bibinfo {volume} {1998}},\ \bibinfo {pages} {1369} (\bibinfo {year} {1998})}\BibitemShut {NoStop}%
\end{thebibliography}%

\end{document}